\documentclass[12pt]{iopart}

\usepackage{cite}
\usepackage{amssymb}
\usepackage{xcolor}
\usepackage{url} 
\usepackage[hidelinks]{hyperref}
\hypersetup{breaklinks=true}

\usepackage{lineno}

\usepackage{graphicx}

\newcommand{\revcolor}{black}
\newcommand{\revcolorN}{black}
\newcommand{\microm}{µm}
\newcommand{\microrad}{µrad}

\newcommand{\angdeg}{^\circ}
\newcommand{\uphotons}{\mathrm{photons}}
\newcommand{\emass}{m_{e}}
\newcommand{\omegaLaser}{\omega_{0}}
\newcommand{\anot}{a_{0}}
\newcommand{\chie}{\chi_{e}}

\newcommand{\SNR}{\mathrm{SNR}}

\usepackage[colorinlistoftodos]{todonotes}

\begin{document}

\title{Single Particle Detection System for Strong-Field QED Experiments}

\author{F~C~Salgado$^{1,2}$, N~Cavanagh$^{3}$, M~Tamburini$^{4}$, D~W~Storey$^{5}$, R~Beyer$^{6}$, P~H~Bucksbaum$^{7}$, Z~Chen$^{5}$, A~Di~Piazza$^{4}$, E~Gerstmayr$^{7}$, Harsh$^{1,2}$, E~Isele$^{7}$, A~R~Junghans$^{6}$, C~H~Keitel$^{4}$, S~Kuschel$^{8}$, C~F~Nielsen$^{9}$, D~A~Reis$^{7}$, C~Roedel$^{10}$, G~Sarri$^{3}$, A~Seidel$^{1,2}$, C~Schneider$^{6}$, U~I~Uggerh{\o}j$^{9}$, J~Wulff$^{1,2}$, V~Yakimenko$^{5}$, C~Zepter$^{1,2}$, S~Meuren$^{7}$, and M~Zepf$^{1,2}$}
\address{$^1$ Institut f{\"u}r Optik und Quantenelektronik, Friedrich-Schiller-Universit{\"a}t Jena, Max-Wien-Platz 1, 07743 Jena, Germany}
\address{$^2$ Helmholtz-Institut Jena, Fr{\"o}belstieg 3, 07743 Jena, Germany}
\address{$^3$ Queen's University Belfast, University Road, BT7 1NN Belfast, United Kingdom}
\address{$^{4}$ Max-Planck-Institut f{\"u}r Kernphysik, Saupfercheckweg 1, D-69117 Heidelberg, Germany}
\address{$^{5}$ SLAC National Accelerator Laboratory, Menlo Park, California 94025, USA}
\address{$^{6}$ Institute for Radiation Physics, Helmholtz-Zentrum Dresden-Rossendorf, Germany}
\address{$^{7}$ Stanford PULSE Institute, SLAC National Accelerator Laboratory, Menlo Park, California 94025, USA}
\address{$^{8}$ Institut f{\"u}r Experimentalphysik, University of Hamburg, Germany}
\address{$^{9}$ Department of Physics and Astronomy, Aarhus University, 8000 Aarhus, Denmark}
\address{$^{10}$ Institute of Nuclear Physics, Technical University Darmstadt, Schlossgartenstr. 9, 64289 Darmstadt, Germany}

\ead{felipe.salgado@uni-jena.de}

\vspace{10pt}
\begin{indented}
\item[]December 2021
\end{indented}

\begin{abstract}
Measuring signatures of strong-field quantum electrodynamics (SF-QED) processes in an intense laser field is an experimental challenge: it requires detectors to be highly sensitive to single electrons and positrons in the presence of the typically very strong x-ray and $\gamma$-photon background levels. In this paper, we describe a particle detector capable of diagnosing single leptons from SF-QED interactions and discuss the background level simulations for the upcoming Experiment-320 at FACET-II (SLAC National Accelerator Laboratory). The single particle detection system described here combines pixelated scintillation LYSO screens and a Cherenkov calorimeter. We detail the performance of the system using simulations and a calibration of the Cherenkov detector at the ELBE accelerator. Single~3~GeV leptons are expected to produce approximately~537 detectable photons in a single calorimeter channel. This signal is compared to Monte-Carlo simulations of the experiment. A signal-to-noise ratio of~{\color{\revcolor}18} in a single Cherenkov calorimeter detector is expected and a spectral resolution of 2\,\% is achieved using the pixelated LYSO screens.
\end{abstract}
%
\vspace{2pc}
\noindent{\it Keywords}: strong-field QED, pair-creation, single-particle detection, Cherenkov calorimeter, Breit-Wheeler process.
%
\submitto{\NJP}
%
\maketitle
%
%

\section{Introduction}

The interaction between light and matter is described by the theoretical framework of quantum electrodynamics (QED). In the perturbative limit, it is considered to be the most precise  and well-tested theory of modern physics~\cite{karshenboim2006study}. As the electric field strength approaches the so-called Schwinger critical field $E_{s} \approx 1.3\times 10^{18} \mathrm{V/m}$, novel strong-field quantum effects become important. Consequently, the description of electron-laser interactions must be described by dressed states as $\anot \gg 1$ and by high-order processes with radiative corrections scaling with $\alpha\chie^{2/3}$ included in the theory~\cite{Ritus.1985}, which is referred to as strong-field QED (SF-QED), where $\alpha$ is the fine-structure constant. 

The normalized vector potential $\anot$ is a Lorentz invariant given, in Heaviside-Lorentz natural units ($c=\hbar=\epsilon_0=1$), by $\anot = eE/(\emass \omegaLaser )$. Here, $e$ is the absolute electron charge, $E$ the peak value of the laser electric field and $\emass$ is the rest mass of the electron~\cite{Ritus.1985}. Another important parameter in the theory of SF-QED is the quantum parameter $\chie$. It is defined, also in natural units, as $\chie = \anot\gamma_e(\omegaLaser/\emass)(1 - \cos\theta)$, where $\gamma_e$ is the colliding electron beam Lorentz-factor, $\omegaLaser$ is the laser frequency, and $\theta$ is the collision angle between the electron with the laser beam (for head-on collision $\theta=180^{\circ}$). A parameter $\chie \gtrsim 1$ indicates that high-energy photon emission by the electron is likely, and, therefore, the particle undergoes a significant recoil on its motion.

Fundamental processes involving photon emission and photon decay are modified under strong-fields and mechanisms such as multi-photon Compton scattering, radiation reaction and nonlinear Breit-Wheeler (BW) are examples of predictions resulting from SF-QED. However, the experimental investigation of this regime is very limited to date~\cite{diPiazza2012}. Few experiments utilize the high-intensity fields in collisions of ultra-relativistic ions~\cite{Aaboud2017} or the interaction of ultra-relativistic particles in aligned crystals\cite{Nielsen2020}. Recently, the investigation of SF-QED has been proposed using the collision of tightly focused ultra-relativistic electron beams~\cite{Yakimenko2019}.

With the advent of ultra-intense laser pulses~\cite{strickland1985compression, Mourou2006}, strong electric fields can be achieved in the laboratory by strongly focusing ultra-intense lasers. However, the Schwinger critical field is still far beyond  the reach of present laser technology by around 3-4 orders of magnitude~\cite{lureau2020high}. A solution for this challenge is to combine the highest achievable electric field from laser pulses with ultra-relativistic electron beams or $\gamma$-photons, allowing the Schwinger field to be achieved in the rest frame of the electrons. In addition to pair generation, the vacuum responds nonlinearly and processes such as light-light scattering~\cite{Gies2018} and vacuum birefringence~\cite{Heinzl2006} can occur and be detected~\cite{Karbstein.2015}. A review of strong-field QED processes is found in Ref.~\cite{diPiazza2012}.

The first experiments in strong-field QED using intense laser fields and ultra-relativistic electron beams were reported in the Experiment-144 at SLAC in the 1990s~\cite{Bula.1996, Burke.1997, Bamber.1999}. In this experiment, electron bunches were accelerated by a linear accelerator up to energies of~49.1~GeV and interacted with a laser field with a root mean square (RMS) normalized vector potential of~$\anot^{RMS} = 0.4$ and $\chie^{RMS} \approx 0.3$ producing about~100 positrons in total in the perturbative multi-photon regime  ($\chie < 1$ and $\anot < 1$)~\cite{Bula.1996, Burke.1997, Bamber.1999}. 

Experiments have been proposed for investigating SF-QED effects based on the interaction of high energy electron or photons beams with high-intensity lasers or nuclear fields in crystals~\cite{Reiss.1971, Nielsen2020}. Moreover, experiments investigating pair production, multi-photon Compton scattering and radiation reaction using all-optical setups have been proposed and realised at the Astra-Gemini laser system at the Rutherford Appleton Laboratory (RAL)~\cite{RALProposal.2010, Cole.2018, Poder.2018}. In such experiments, electron beams generated by laser-wakefield acceleration (LWFA) up to~2~GeV interacted with an intense laser pulse with~$\anot = 10$~\cite{Cole.2018, Poder.2018} at the nonperturbative moderate quantum regime ($\chie < 1$ and $\anot > 1$), and signatures of radiation reaction process were observed. Upcoming projects aim to investigate SF-QED effects in the nonperturbative full quantum regime, i.e. $\chie > 1$ and $\anot > 1$. Interactions can be between two beams of photons \cite{Gies2018} or between electron beams and intense laser pulses as proposed in the Experiment-320 (E-320) at FACET-II~\cite{meuren2019probing, meuren2020probing, WorkshopSFQED.2019} and the LUXE experiment at DESY~\cite{Hartin.2019, luxe_letter.2019, LUXECDR.2021}. In both experiments, a small number of electron-positron pairs, which are generated by SF-QED processes must be measured with a sensitive detection system. The challenge in the detector development is that the detection system must be able to detect single particles, while also being insensitive to the photon background which is inherent to the experiments with an ultra-relativistic electron beam and a beam dump close to the interaction region. 

\begin{figure}[h!]
    \centering
    \includegraphics[width=0.7\textwidth]{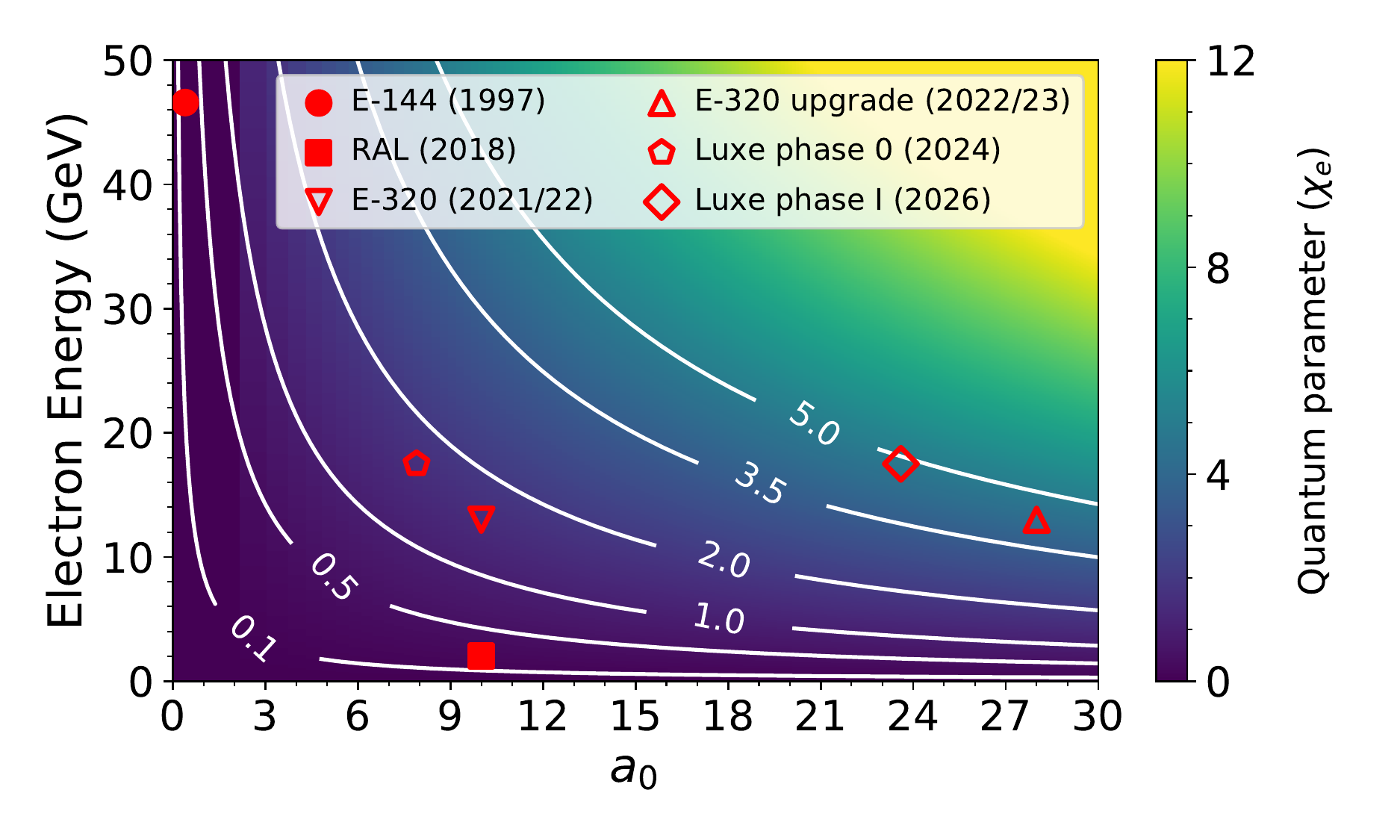}
    \caption{Electron quantum parameter $\chie$ for different normalized laser strength $\anot$ and incoming electron energies. The filled red markers indicates the already performed experiments: (circle) Experiment-144 (E-144), 1997~\cite{Bula.1996, Burke.1997}; (square) RAL, 2018~\cite{Poder.2018}. The open markers shows the upcoming SF-QED experiments: (up-pointing triangle) Experiment-320 (E-320), 2021/2022~\cite{meuren2019probing}; (down-pointing triangle) Experiment-320 upgrade, 2022/2023~\cite{meuren2019probing}; (pentagon) LUXE phase 0, 2024~\cite{LUXECDR.2021}; (diamond) LUXE phase 1, 2026~\cite{LUXECDR.2021}. The Experiment-320 will probe the nonperturbative full quantum regime of interaction ($\chie > 1$ and $\anot > 1$).}
    \label{fig. a0 map}
\end{figure}

In \Fref{fig. a0 map}, the different regimes of interaction of each SF-QED experiment with electron-laser interaction are highlighted.

Here we describe the detection system which is designed for SF-QED experiments such as the E-320 at FACET. We report a detection system which is able to diagnose single particles with MeV to GeV energies. A test and calibration of the detector system is presented. Monte-Carlo simulations of the background noise level and the expected signal-to-noise ratio for the E-320 show that single positron events can be detected with an expected {\color{\revcolorN} signal-to-noise ratio of 18}.

\section{Single Particle Detection System}

\begin{figure}[htp]
    \centering
    {\small\hspace{-10cm}a)}\par\includegraphics[width=0.85\linewidth]{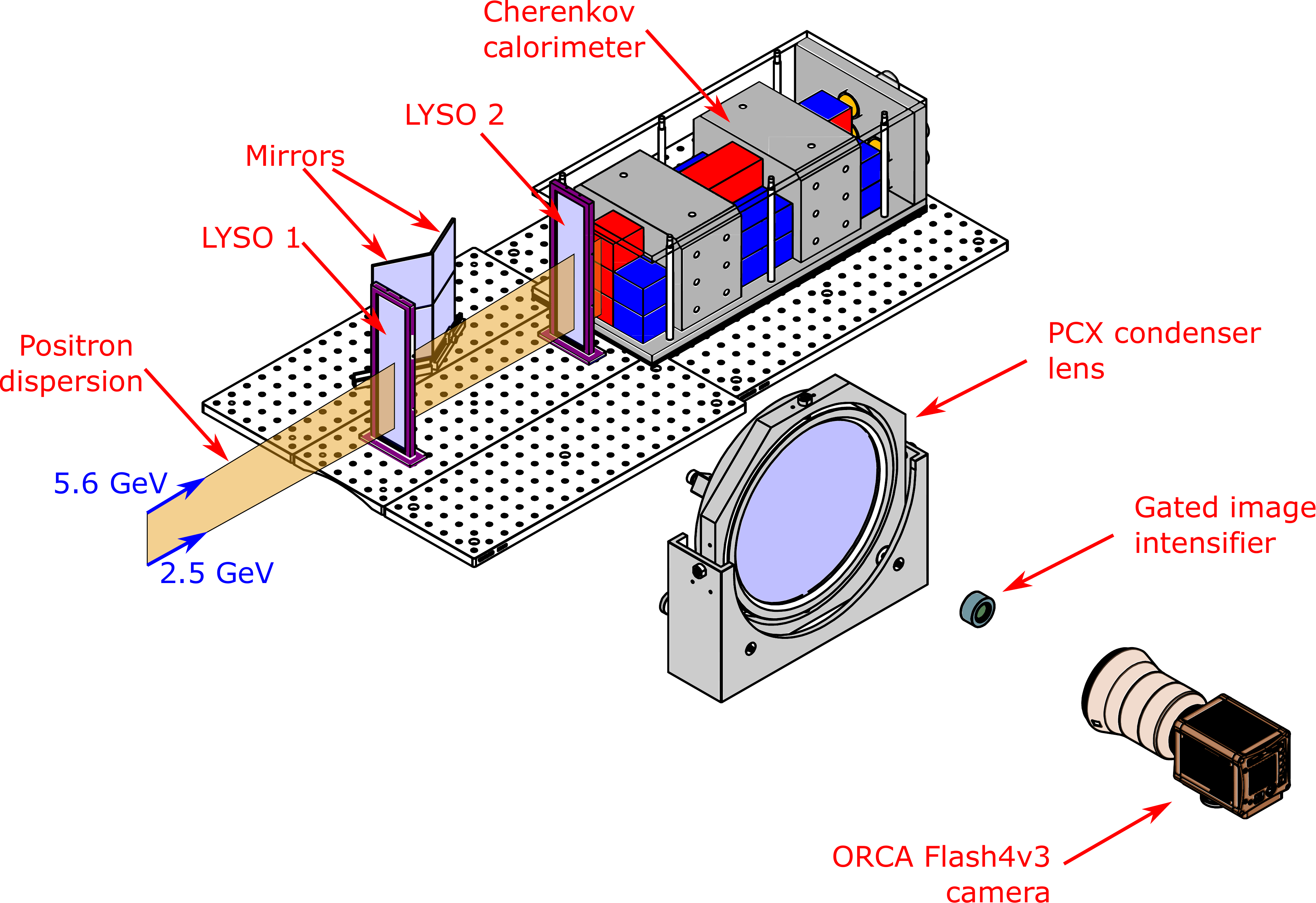}\par
    {\small\hspace{-10cm}b)} \par
    \includegraphics[width=0.6\linewidth]{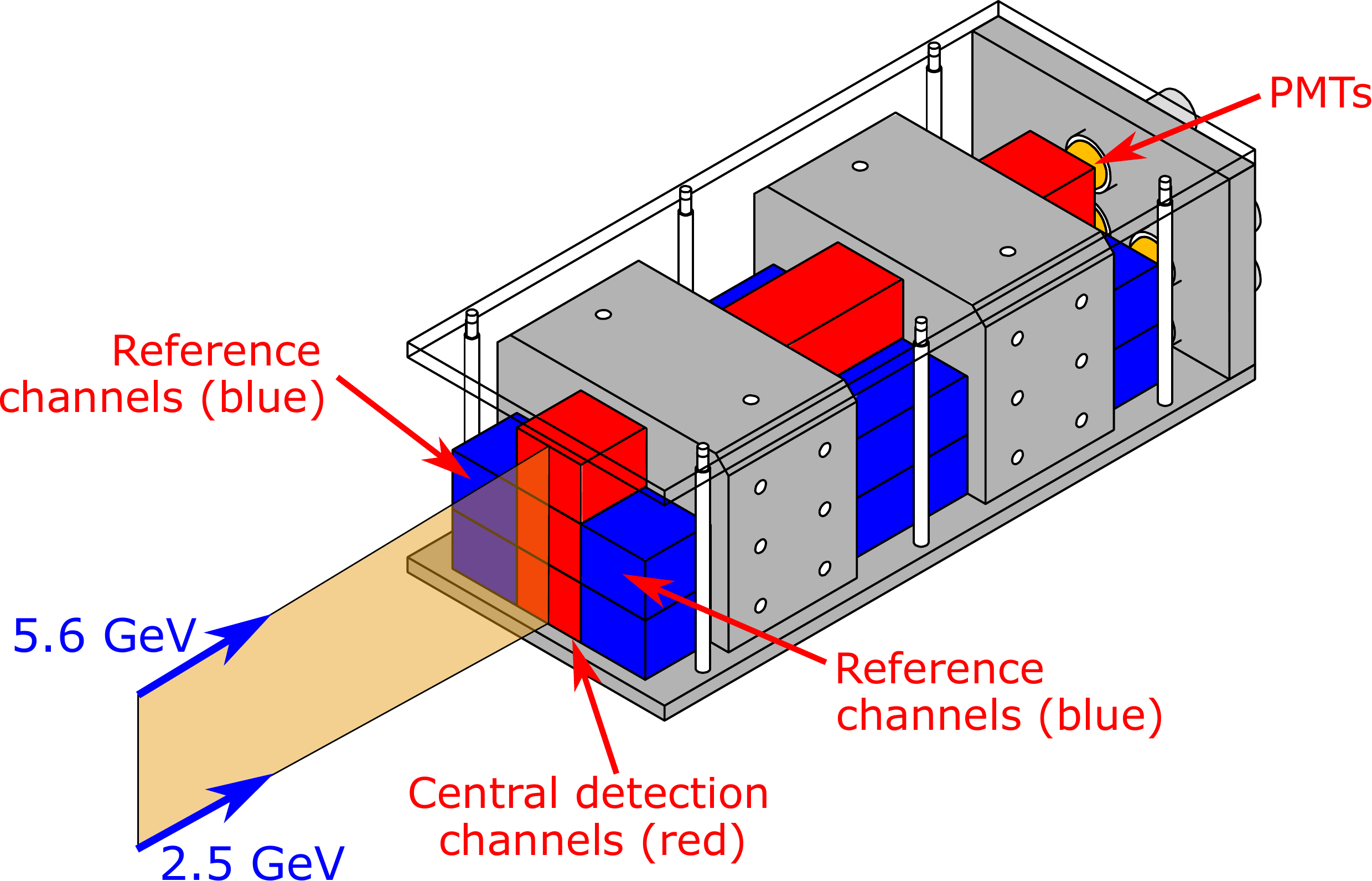}
    \caption{a) Proposed design of the single particle detection system {\color{\revcolor}(not to scale)} for E-320 at FACET-II. The incident single~GeV-positron travels through two pixelated LYSO scintillating screens which provides particle tracking and high resolution spectral information before entering the Cherenkov detector at one of its lead-glass detection channels where its energy is fully deposited. b) Detailed view of the Cherenkov detector. The background reference channels, where no signal particle is expected to strike, are shown in blue color. The central detection channel of the detector, which signal leptons are deflected, are illustrated in red color. The Cherenkov photons produced inside the lead-glass channel are detected by photomultiplier tubes (PMTs) at the rear of each channel. The positron direction of dispersion is within the yellow area with limits of 2.5~-~5.6~GeV.}
    \label{fig. proposed single particle systems}
\end{figure}

The detection system comprises two pixelated LYSO:Ce crystal screens to provide high spatial resolution coupled to a segmented Cherenkov calorimeter both placed about~3.6~m after a dipole magnet. The screens are placed in front of a Cherenkov calorimeter, and thus provide the ability to discriminate against low energy background events. Both detectors have fast (nanosecond-scale) response which allows timing-based background suppression as well. \Fref{fig. proposed single particle systems}a) presents the setup of the single particle detection system. 

The two pixelated LYSO screens have dimensions of 4~cm $ \times$ 20~cm $\times$ 4~mm with crystal pixels sizes of 2~mm $\times$ 2~mm $\times$ 4~mm which provide tracking information on the single particles propagating towards the detector. The LYSO:Ce crystal has a scintillation yield of 25~photons/keV emitted at central wavelength of 410~nm~\cite{AdvatechLYSO}. Besides the high light yield, the crystal has a decay time of about~40~ns which allows the background noise from secondary sources of radiation located far from the detectors to be suppressed by imaging the screens using a setup combination of a condenser lens, a gated image intensifier and single-photon sensitive camera Hamamatsu ORCA-Flash4~\cite{Photek.2021, Orcas.2021}, see calibration of the LYSO crystals in~\ref{sec: LYSO:Ce decay time calibration}. A PCX condenser lens (diameter of 25~cm and focal length of 40~cm) is placed about 800~mm away from the screens and the image intensifier placed about 270~mm after the lens giving enough magnification to image the entire active region of the LYSO screens at the image intensifier. The Hamamatsu ORCA camera is placed about 250~mm away from the image intensifier with a f/1.4 macro lens with focal length of 28~mm and 40~mm diameter to maximise the photon collection.

\begin{figure}[htp]
    \centering
    \includegraphics[width=0.95\linewidth]{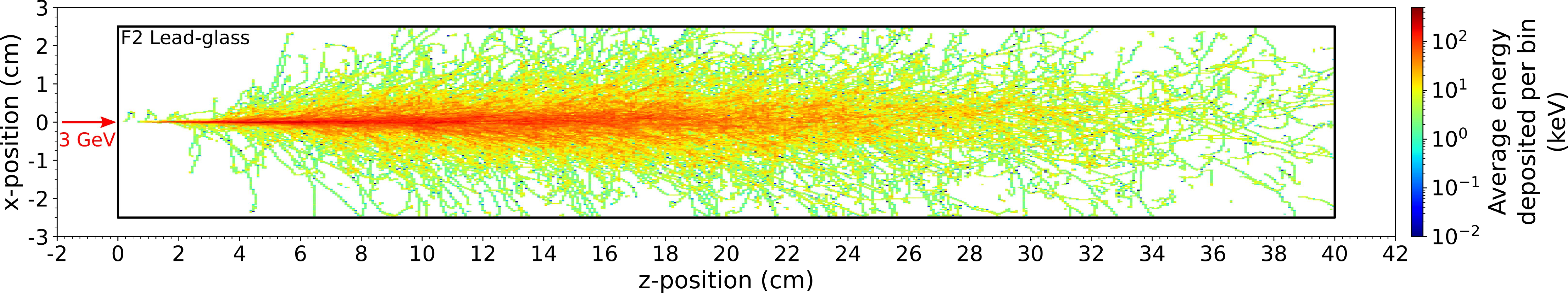}
    \caption{Simulated energy deposition for a single 3 GeV positron inside a single F2 lead-glass block 50~mm $\times$ 40~mm $\times$ 400~mm of the Cherenkov calorimeter. The lateral electromagnetic shower is well contained within the Moli{\`e}re radius ($R_M=\mathrm{3.4~cm}$) inside a single calorimeter block. Each bin has an area of~$\mathrm{0.1~mm}^2$.}
    \label{fig. Dosedep}
\end{figure}

The Cherenkov calorimeter, which is placed behind the pixelated LYSO screens, comprises up to~7~detection channels of 50~mm $\times$ 40~mm $\times$ 400~mm Schott F2~lead-glass wrapped in enhanced specular reflector (ESR) foils to prevent the optical photon cross-talk between the detection channels. The choice of F2 lead-glass for the Cherenkov calorimeter is due to its linear response for energy measurements between 1 - 4~GeV and absence of scintillation as discussed in~\cite{Bartoszek.1991, Kirsebom.1986}. The F2 lead-glass has a radiation length of~$X_0=$ 3.14~cm and a Moli{\`e}re radius of $R_M=$ 3.4~cm. Hence, the glass blocks of the Cherenkov calorimeter were designed to contain the particle shower produced by a single 3~GeV positron incident on it as shown in~\Fref{fig. Dosedep}. The calorimeter array consists of up to~7~blocks with the signal positrons designed to be incident on the the 3~central channels, which corresponds to a total active area {\color{\revcolorN} of $3 \cdot (\mathrm{40~mm} \cdot \mathrm{50~mm}) = \mathrm{6000~mm}^2$}, and positioned to detect positrons initially in the 2.5~-~5.6~GeV range for a nominal~87.2~MeV transverse kick, equivalent to an integrated field strength of $\mathrm{BL} = \mathrm{0.3~Tm}$, of the dipole magnet. However, the kick settings of the dipole magnet can be selected for detection of particles at alternative energy ranges. The side channels provide on shot background reference to allow better discrimination of signal events. Detection of the Cherenkov photons is achieved with photomultiplier tubes (PMTs) placed at the rear of each detection channel~\cite{PMTdatasheet}. The PMTs used on the detector have a rise time of about~3~ns therefore also allowing background noise rejection by temporal gating. A detailed view of the Cherenkov detector is shown in \Fref{fig. proposed single particle systems}b) where the reference background channels are colored in blue, the main three central detection channels are in red, and the dispersion direction of the positrons within the 2.5~-~5.6~GeV range is represented by the yellow area. The positron spectrum from the SF-QED interaction is discussed later in \Sref{sec. proposed implementation for the e320 experiment}.

\subsection{Calibration of the Cherenkov Calorimeter}
\label{sec. calibration of the Cherenkov calorimeter}

The Cherenkov calorimeter was calibrated at the ELBE radiation source at the Helmholtz-Zentrum Dresden-Rossendorf (HZDR) using the dark current of the accelerator, which provides single electrons of 27~MeV energy with {\color{\revcolor}a weighted average number of ($0.156 \pm 0.005$)~electrons/RF cycle}. The calibration of the ELBE accelerator dark current is presented in detail in~\ref{sec: Dark Current Calibration}.

\begin{figure}[htp]
    \centering
    \includegraphics[width=0.45\linewidth]{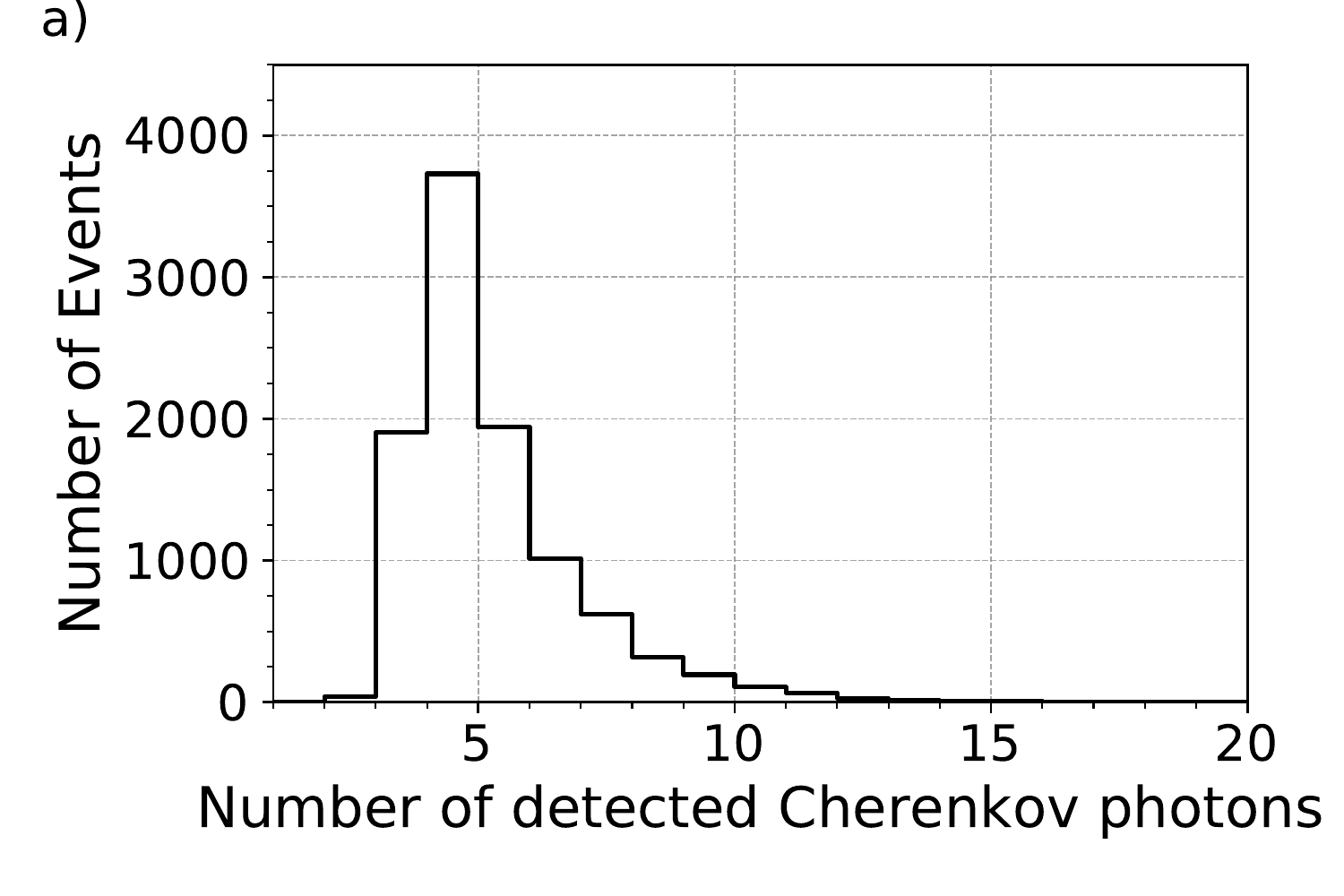}\quad
    \includegraphics[width=0.45\linewidth]{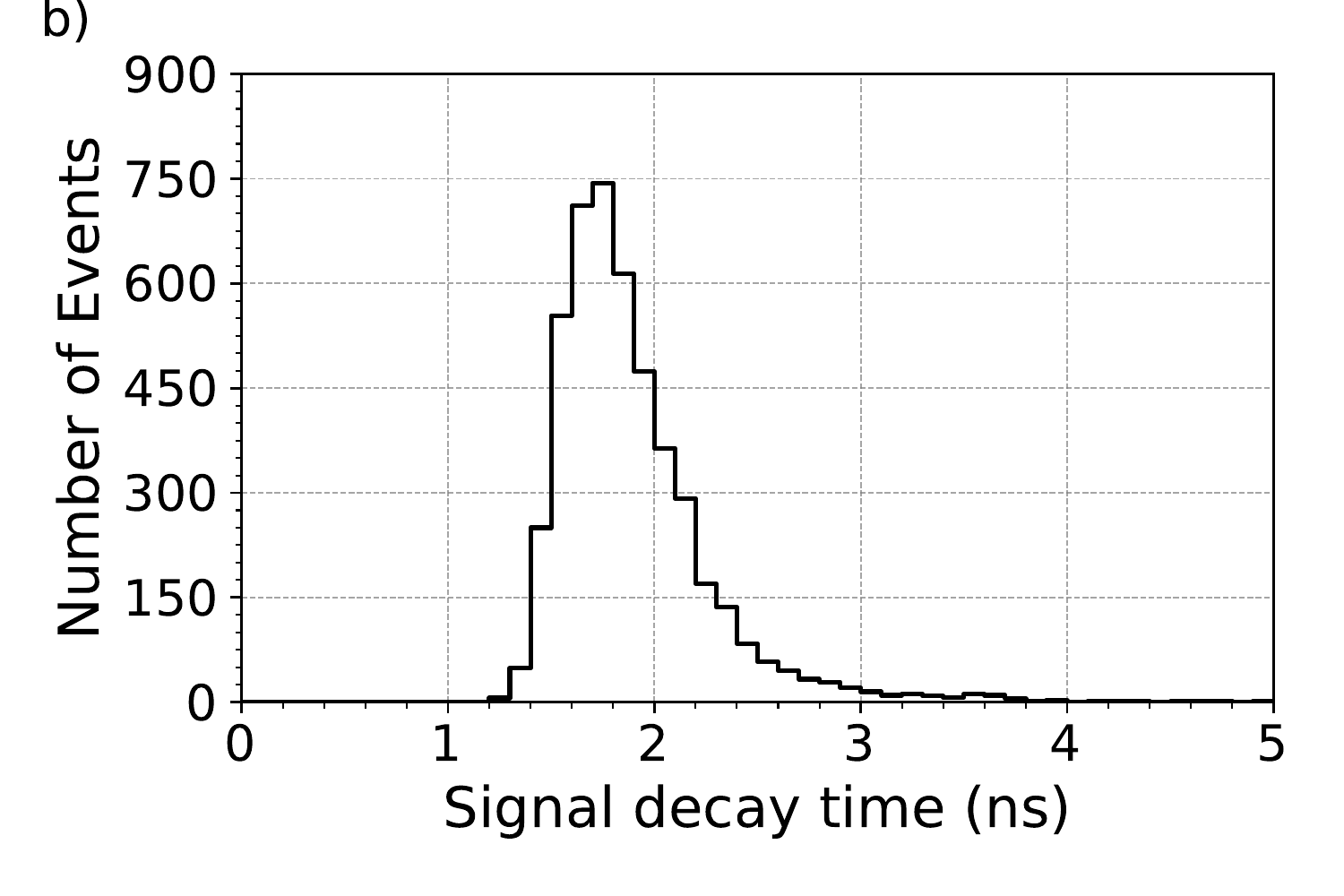}
    \caption{Cherenkov calorimeter calibration results using the single electrons obtained using the dark current of ELBE radiation source at the Helmholtz-Zentrum Dresden-Rossendorf (HZDR). The dark current of the linear accelerator is composed of single~27~MeV electrons. a) Number of Cherenkov photons detected by the Cherenkov calorimeter - an average of~4~photons per single electron is detected; b) decay time of the recorded signal - a single electron hit produces a signal which is about~1.8~ns long.}
    \label{fig. Cherenkov Calorimeter calibration}
\end{figure}

The dark current measurement was performed on the central calorimeter channel with the PMT gain set to~$4 \times 10^6$, corresponding to a voltage across the cathode and anode of the PMT of~$10^3$~V. The signals of the PMTs were recorded using PicoScopes~\cite{Picoscope.2020}. A total of~$10^4$ events were acquired, and the number of Cherenkov photons detected by the PMTs as well as the signal decay time of each event are shown in~\Fref{fig. Cherenkov Calorimeter calibration}. 

\begin{figure}[htp]
    \centering
    \includegraphics[width=0.5\linewidth]{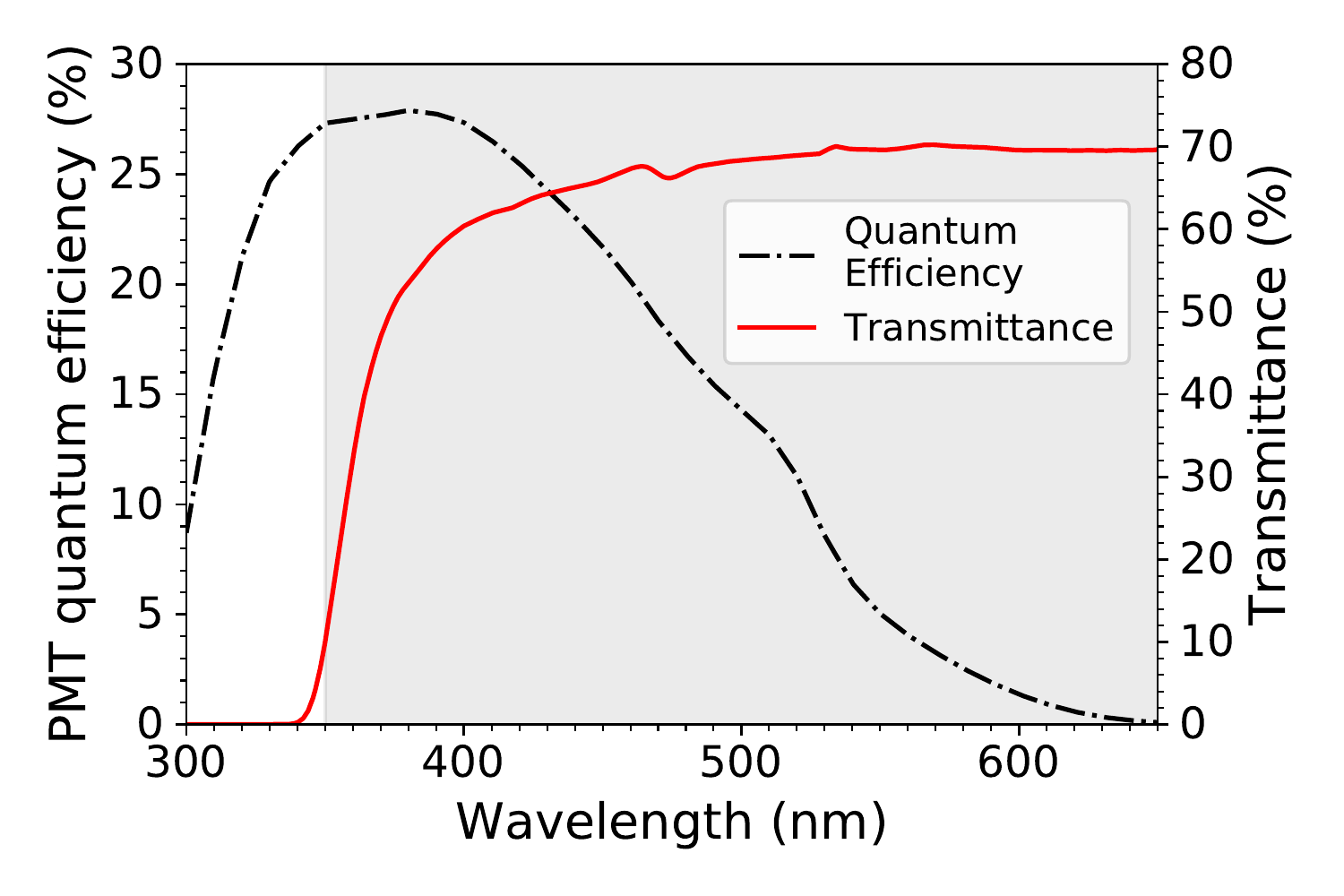}
    \caption{Typical quantum efficiency of the photomultiplier tube (PMT) employed on the Cherenkov calorimeter and the calibrated transmittance of the 40~cm lead-glass. The gray shaded region shows the wavelength region between 350~-~650~nm where both parameters were implemented in the GEANT4 model of the detector to evaluate the number of photons hits on the PMT.
    }
    \label{fig. PMT QE and lead-glass transmittance}
\end{figure}

\begin{figure}[htp]
    \centering
    \includegraphics[width=0.45\linewidth]{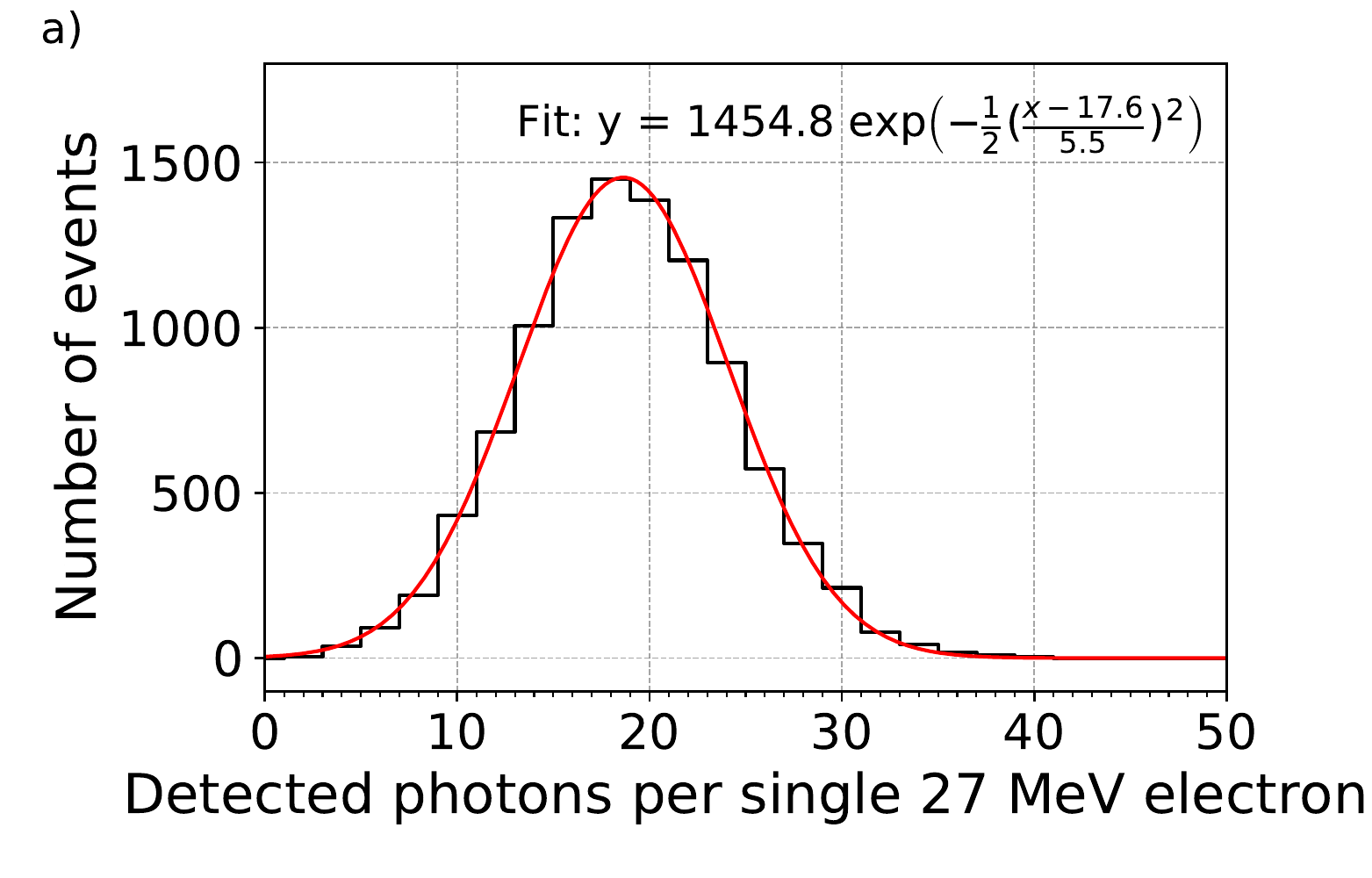}\quad
    \includegraphics[width=0.45\linewidth]{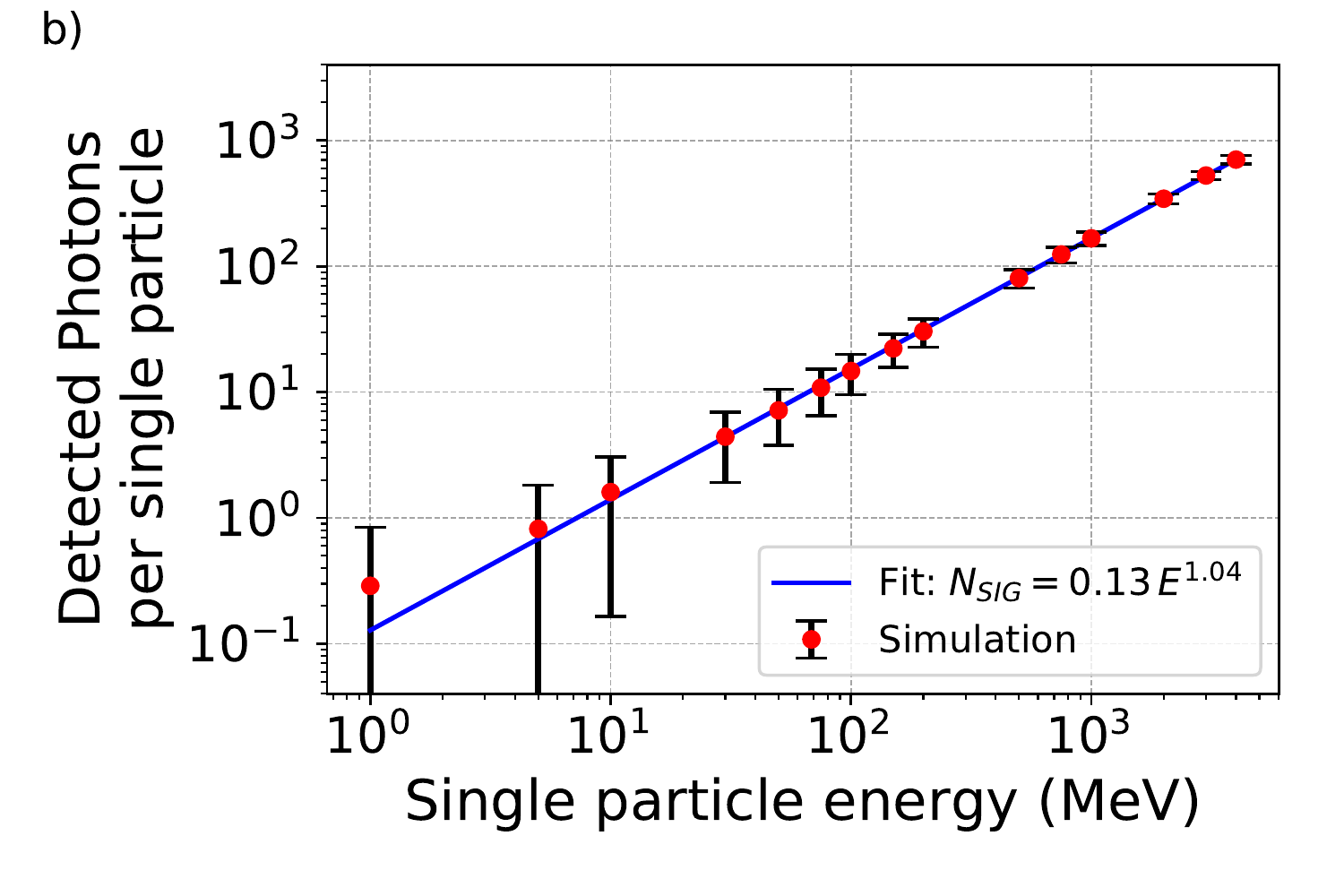}
    \caption{\color{\revcolor}(a) Simulated number of photons detected for a single 27~MeV particle incident centrally on a Cherenkov calorimeter channel. About 17~photons are detected with an RMS variance of 5.5~photons. (b) Calculated number of Cherenkov photons detected per channel for a single incident particle of different energies after the detection efficiency $\eta$ of 23\%  and its uncertainty were taken into account. The error bars indicate the standard deviation which determines the energy resolution of the calorimeter. A single 3~GeV particle results in 537~detected photons by the calorimeter. Hence, a single~GeV-particle incident on the detector is easily detected by the calorimeter.}
    \label{fig. Calibration detected photons at calorimeter channel}
\end{figure}

From~\Fref{fig. Cherenkov Calorimeter calibration}a), an average of~4~photons are detected by the photomultiplier tube after a single electron with~27~MeV energy hits the detection channel. From Monte-Carlo simulations using GEANT4~\cite{agostinelli_geant4simulation_2003, allison_geant4_2006, allison_recent_2016}, it is predicted that a single~27~MeV electron produces~1770~photons of which only 17~photons are detected by the PMT. {\color{\revcolor} \Fref{fig. Calibration detected photons at calorimeter channel}a) shows the simulated distribution using GEANT4 of the number of detected photons by the calorimeter channel. The distribution was fitted by a Gaussian curve with mean at $N_{Sim} = 17.6$~photons and RMS width of $\sigma_{Sim} = 5.5$~photons.} In the simulated model of the Cherenkov calorimeter detector, the transmittance of the lead-glass and the PMT quantum efficiency were included and the generation of optical photons in the range between 350~-~650~nm was simulated, see \Fref{fig. PMT QE and lead-glass transmittance} for the lead-glass transmittance characterised over its 40~cm length and the typical quantum efficiency of the PMT employed. By comparing the simulation obtained for a single 27~MeV hit and the calibration results for an incident particle of 27~MeV, we calculate a photon detection efficiency of {\color{\revcolor}$\eta \pm \sigma_{\eta} = (0.23 \pm 0.13)$, where $\eta = 4/17.6 \approx 0.23$ and $\sigma_{\eta}$ is calculated using error propagation method such that $\sigma_{\eta} = \eta \cdot ( ( \sqrt{4}/4 )^2 + (5.5/17.6)^2 )^{1/2} = 0.13$.}

Based on this calibration and the scaling of Cherenkov photons derived from GEANT4 simulations we calculate the number of detected photons for different energies of incoming single particles, see \Fref{fig. Calibration detected photons at calorimeter channel}b). Approximately~537~photons are detected for a single~3~GeV particle, providing an easily observed signal for single~GeV-particle hits. The spectral resolution of the calorimeter depends on the energy of the incident particle and is limited by statistics of the detected Cherenkov photons. A representative value is  approximately~20\% with a resolution of better than 10\% possible at the highest energy range. Higher spectral resolution and rejection of false positive events is provided by the LYSO tracking screens in combination with the Cherenkov detectors for event rejection as discussed in the following section.

{\color{\revcolor}To guard against any unknown non-linearities, an in-situ calibration of the detection system is planned during the beamtime of the Experiment-E320 using thin foils to produce a few positron-electron pairs with energies in the 2.5~-~5.6~GeV range, and a correction on the calibration curve can be applied during the experiment.}

\section{Proposed Implementation for the Experiment-320: Expected Performance}
\label{sec. proposed implementation for the e320 experiment}

In this section, we present the experimental parameters of the Experiment-320 followed by a discussion of the background noise and the signal-to-noise ratio expected on the single particle detection system.   

The E-320 uses the FACET-II linear accelerator to generate electron beams up to~13~GeV with charge of~2~nC, a maximum Gaussian transverse profile of $\sigma_{e} = 30$~\microm~and divergence less than 6~\microrad~\cite{FACETII.2019}. A flat-top laser beam with diameter of 40~mm is focused by an off-axis parabolic mirror (OAP) reaching a peak intensity of $1.3\times 10^{20}~\mathrm{W/cm}^2$ which corresponds to an $\anot = 10$. The focused laser beam forms a crossing-angle with the electron beam of about 30$\angdeg$ spatially overlapping with only 1\% of the electron bunch, and, consequently, a quantum parameter~$\chie = 1.5$ becomes experimentally accessible~\cite{meuren2019probing, meuren2020probing}. As a result, E-320 will probe a regime where the interaction with the laser is nonperturbative and electron-positron pair-creation occurs in the tunneling regime~\cite{meuren_semiclassical_2016}. 

\begin{figure}[htp]
    \centering
    \includegraphics[width=0.45\linewidth]{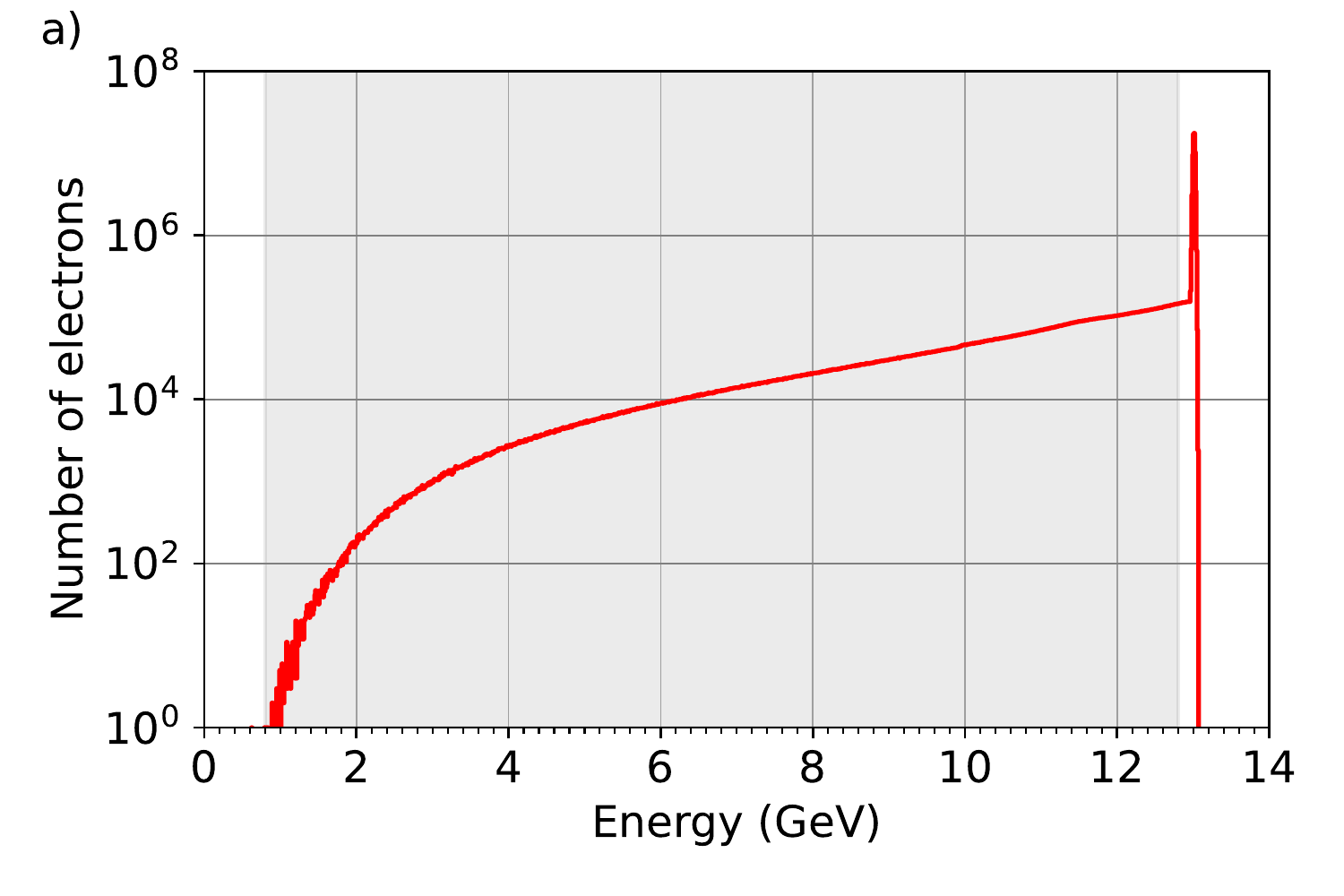}
    \includegraphics[width=0.45\linewidth]{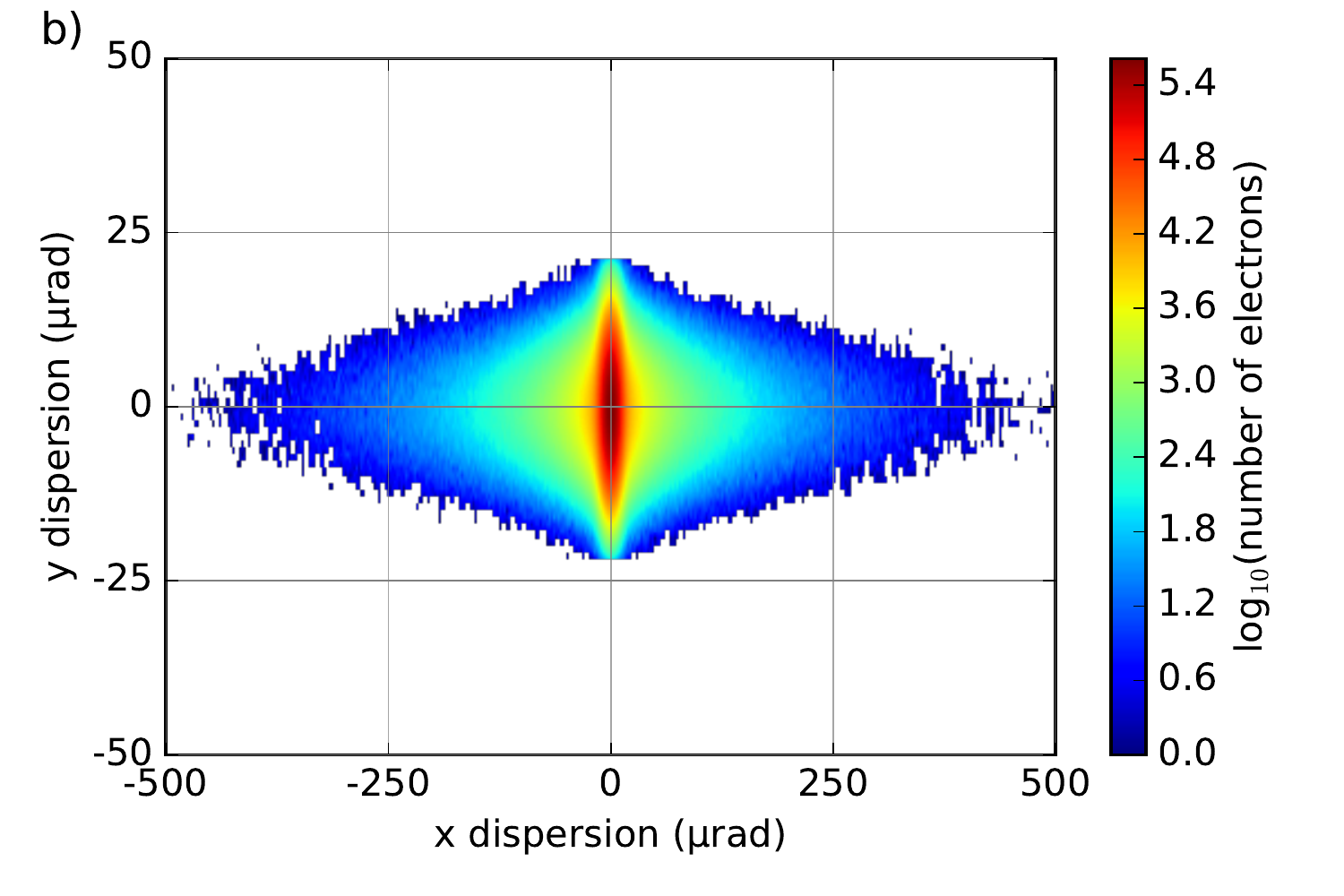}
    \caption{Electron beam a) spectrum (10 MeV bin size) and b) divergence after the interaction with the laser beam. The gray area on the electron spectrum corresponds to the energy range of 1~-~12.8~GeV used on the FLUKA simulations for estimating the background. The FWHM divergence of the electron bunch increases to a maximum of 25~\microrad~after the interaction.}
    \label{fig. electron spectrum and divergence}
\end{figure}

The emission of high-energy photons in the strong laser field {\color{\revcolor} widens the initial monoenergetic 13~GeV electron energy to the range of 1-13~GeV} after the SF-QED interaction and also increases the electron beam divergence to 25~\microrad. The simulated energy spectrum and divergence of the electron beam after the interaction point are shown in \Fref{fig. electron spectrum and divergence} (note that the simulation parameters represent the concept design phase of E320 and final experimental parameters will likely differ from those considered here).

\begin{figure}[htp]
    \centering
    \includegraphics[width=0.45\linewidth]{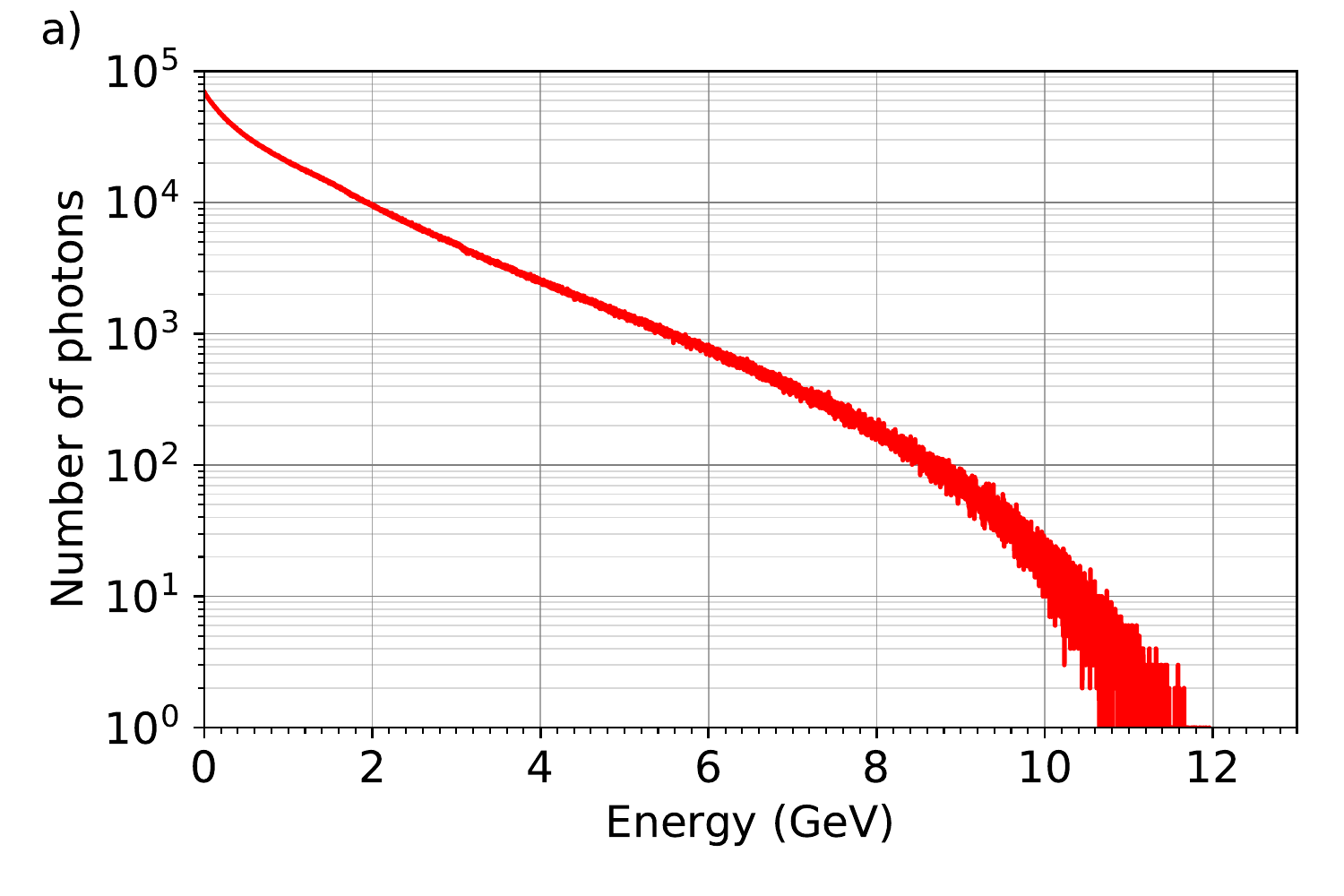}
    \includegraphics[width=0.45\linewidth]{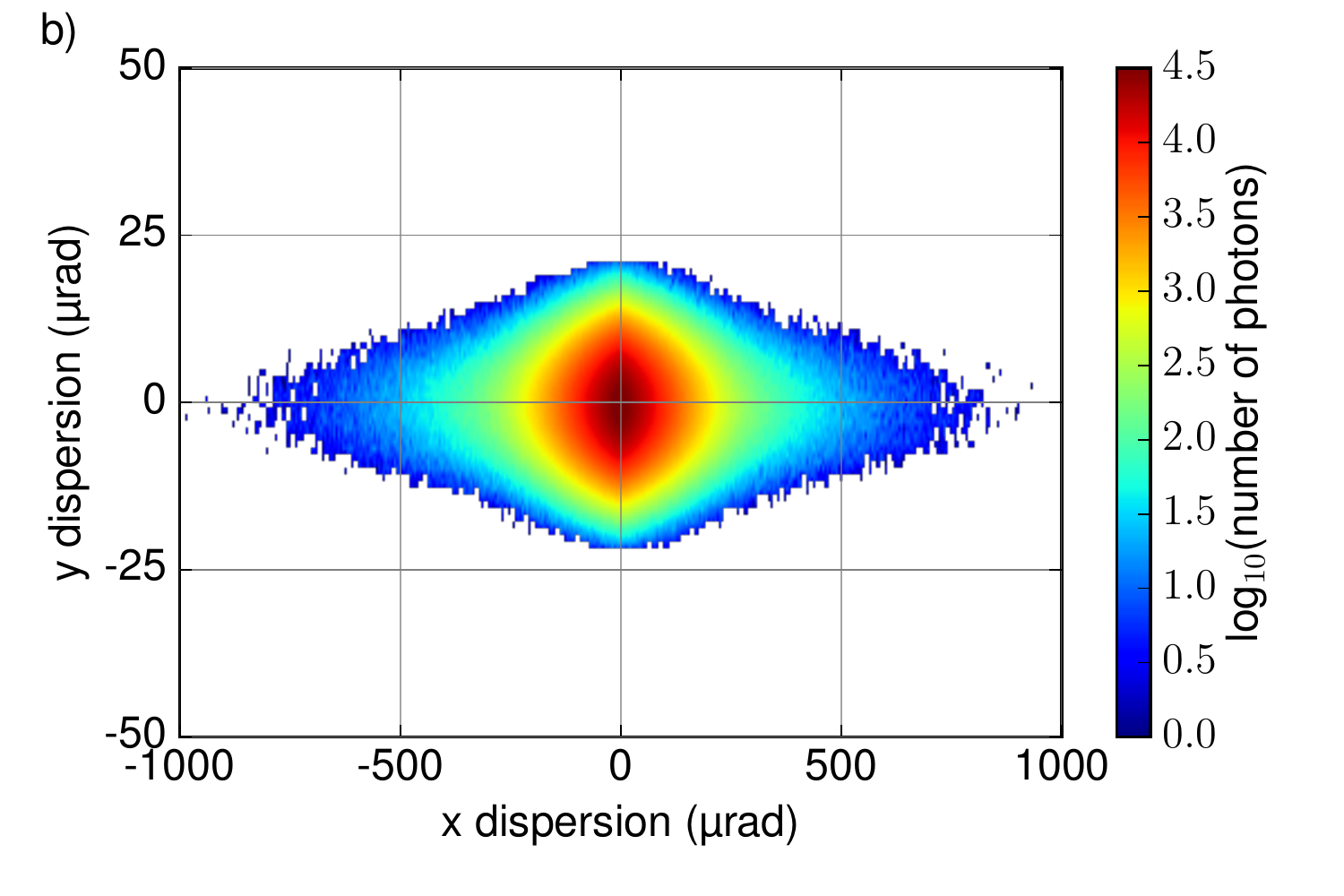}
    \caption{Photon beam a) spectrum (10 MeV bin size) and b) divergence after the electron-laser interaction. The maximum energy reached by the photon beam is limited by the energy of the initial 13~GeV of the primary electrons. {\color{\revcolor} The FWHM divergence of the beam in the x-direction is 134.35~\microrad~since it is the same direction of the collider laser linear polarization. On the other hand, the FWHM divergence in the y-direction is much smaller, about 0.64~\microrad.}}
    \label{fig. photon spectrum and divergence}
\end{figure}

The emitted high-energy photon beam has a maximum energy limited by the electron bunch energy of 13~GeV before the interaction and most of the photons emitted have energies less than 2~GeV. \Fref{fig. photon spectrum and divergence} presents the simulated spectrum and divergence of the photon beam after the electron-laser interaction. Due to the low divergences of the electron and photon beams, neither of them generate substantial background noise at the edge of the OAP or in any other location along the beamline due to their small divergence and short distance to the IP.

\begin{figure}[htp]
    \centering
    \includegraphics[width=0.45\linewidth]{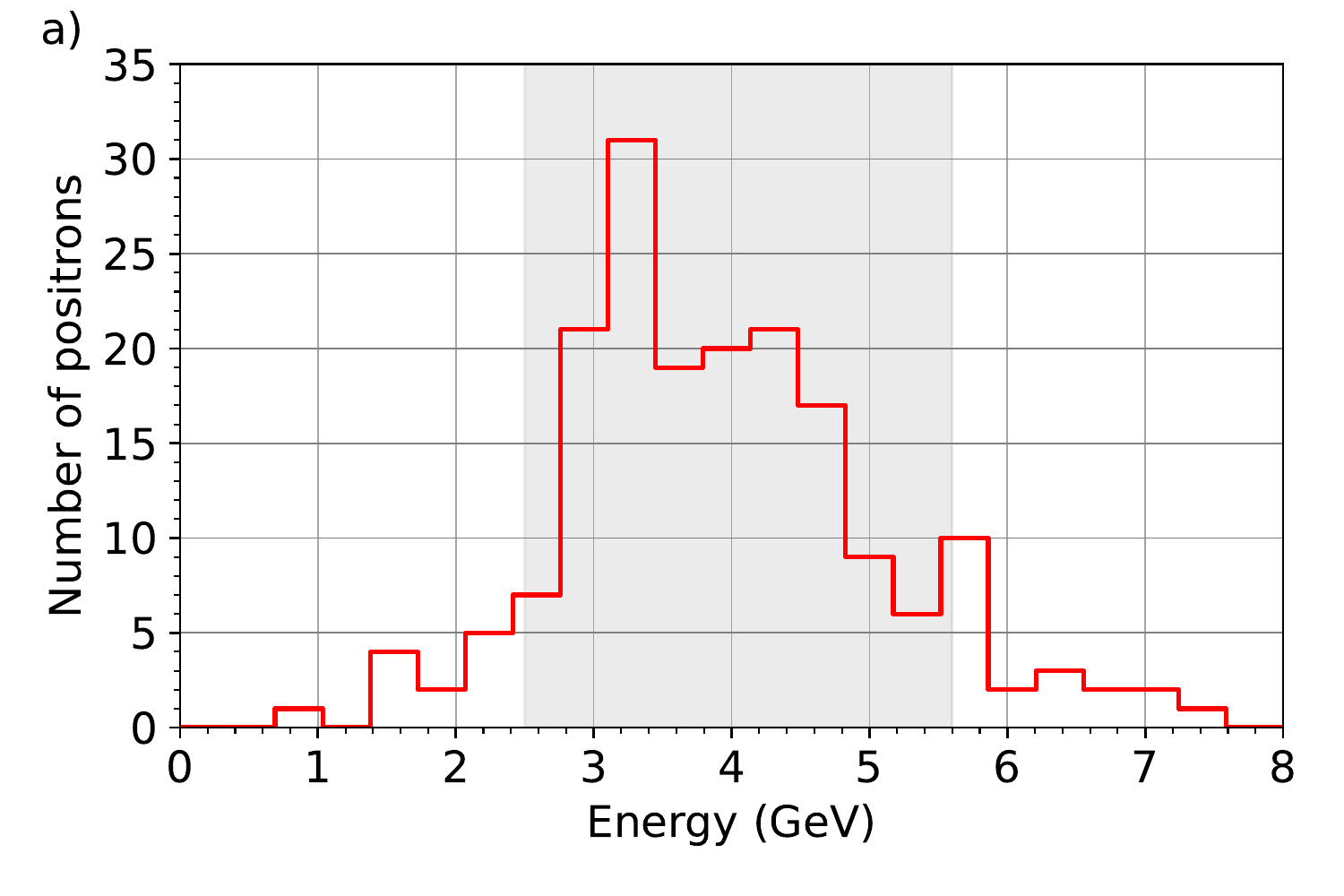}
    \includegraphics[width=0.45\linewidth]{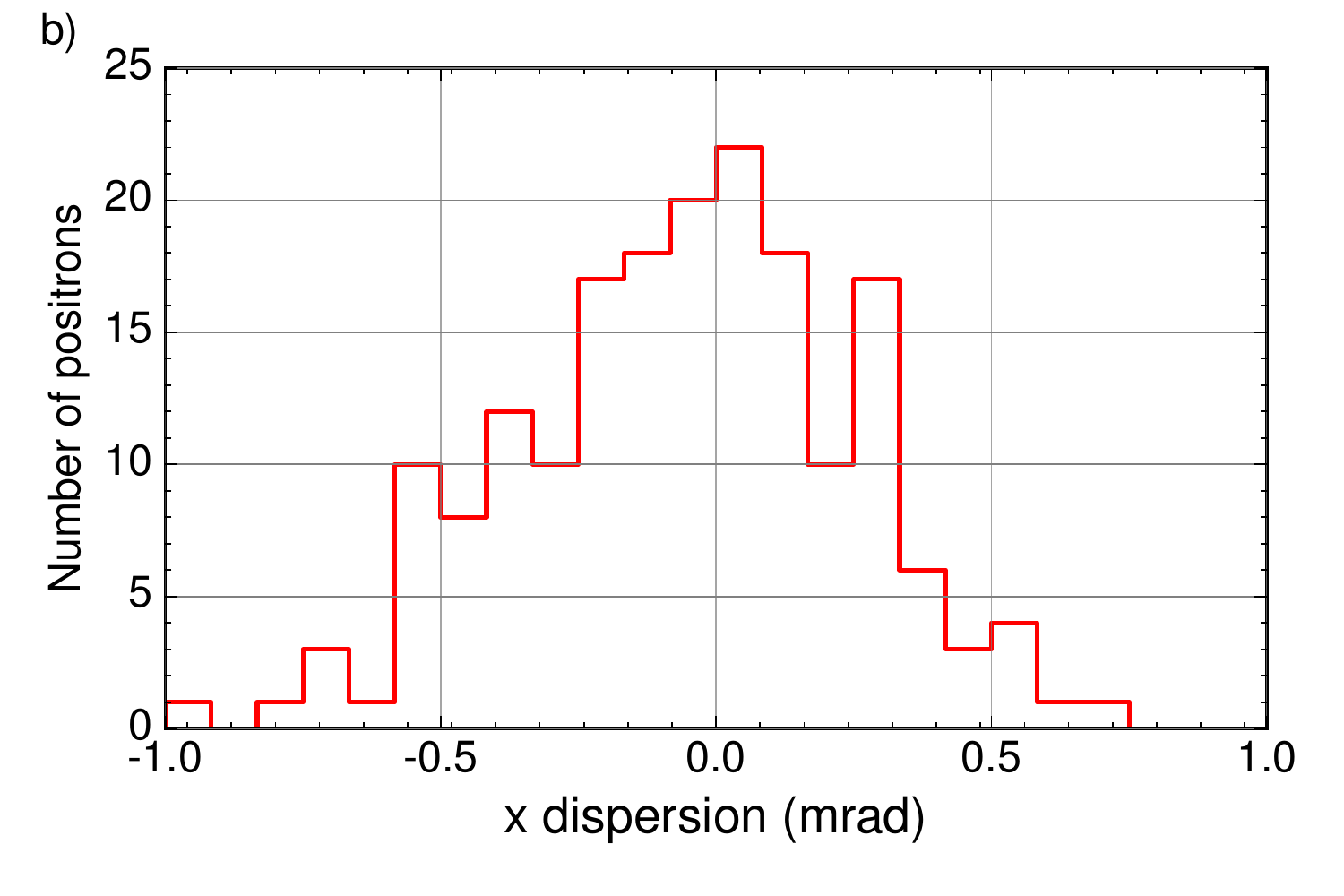}
    \caption{Positron beam a) spectrum (345 MeV bin size) and b) divergence after the interaction with the laser beam. The gray area on the energy spectrum represents the detectable energy range by the detectors between 2.5~-~5.6~GeV with the selected magnet kick of 87.2~MeV.}
    \label{fig. positron spectrum and divergence}
\end{figure}

Some of the emitted high-energy photons, while still immersed in the laser field, interact with the optical photons of the laser and generate electron-positron pairs - predominantly through the nonlinear BW process. The simulation results show that positrons in the range of 1~-~9~GeV have a higher probability of being created as presented in \Fref{fig. positron spectrum and divergence}a), and the particles within the energy range of 2.5~-~5.6~GeV are deflected towards the detectors for the selected magnet kick of 87.2~MeV.

The created pairs propagate through the FACET-II spectrometer beamline alongside with the remains of the primary 13~GeV~electron beam and the high energy photons, up to the dipole magnet where the charged particles are deflected towards the detectors and beam dump. On the other hand, the high-energy photons propagate towards the beam dump without deflection and interaction with any material along its path due to the small divergence. Therefore, no forward background noise contribution on the detectors is expected from the photon beam.

\begin{figure}[htp]
    \centering
    \includegraphics[width=\linewidth, trim={0, 0, 0, 0}, clip]{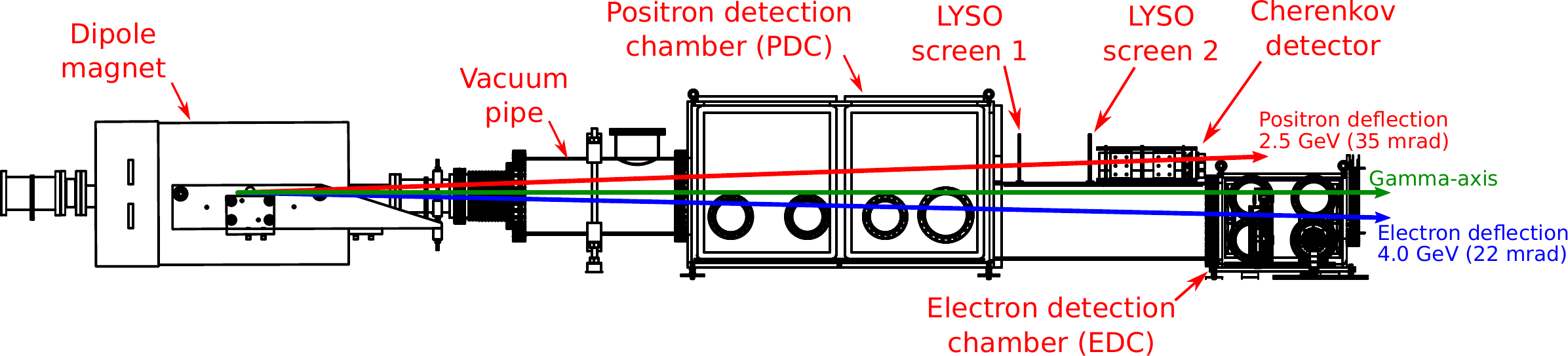}
    \caption{FACET-II beamline with the envisioned position of the single-particle detection system. The dipole magnet disperse the positrons upwards (red trajectory corresponds to a 2.5~GeV positron) where the single particle detection system is installed, and the electrons downwards (blue trajectory corresponds to a 4.0~GeV electron). The gamma beam is represented by the green trajectory.}
    \label{fig. FACET detecting region}
\end{figure}

A slice of the FACET beamline with the positioning of devices such as the dipole magnet and the single particle detector is presented in~\Fref{fig. FACET detecting region}.

The positrons are dispersed upwards by the dipole magnet and they propagate through the positron detection chamber (PDC) before exiting through a~5~mm thick aluminum vacuum exit window to reach the LYSO pixelated screens, and, finally, the Cherenkov detector. On the other hand, electrons are dispersed downwards with particles deflected in the 22~-~65~mrad range being detectable by detectors in the electron detection chamber (EDC), while electrons deflected at angles greater than 65~mrad strike the chamber floor and beamline support structures and thus becoming a background source for the detectors.

A beam dump is placed approximately~8.6~m downstream from the Cherenkov detector to stop the main electron beam and the produced high-energy photons. When the high-energy particles interact with the beam dump, substantial radiation in the backward direction that has the potential to reach the detectors and becomes noise is generated. However, the large distance between the detectors and the beam dump corresponds to a delay of~57~ns between a particle signal hit and the backscattered radiation on the detectors which is enough to significantly reduce dump noise at the detectors by time gating the PMTs and LYSO screens. Hence, the remaining background at the LYSO screens and detectors considered in the simulations below originates from the upstream radiation sources and secondary particles sources located very closely to the detectors which arrive within the gating window of the detectors.

\section{Background and Signal-to-Noise Ratio Estimates for the Experiment-320}

To efficiently detect single particles with high confidence, the design goal for the Cherenkov detectors was to achieve a signal-to-noise ratio in terms of signal power to background power ($\mathrm{SNR}_P=P_S/P_{BG}$) approximately equal or greater than unity and a signal-to-noise ratio in terms of mean to variance ($\SNR_\sigma=\mu/\sigma \gg 1$, where $\mu$ and $\sigma$ are mean value of signal and variance of the background respectively). This is particularly important for data taking at low signal rates, where the number of pairs to be detected per shot is $\approx 1$ or less. Under these circumstances the Cherenkov detector allow the rejection of false positive events in the tracking detector and therefore the accumulation of tracking data to record high resolution spectra at low event rates.

Monte-Carlo simulations were performed using the code FLUKA~\cite{fluka_2005, bohlen_fluka_2014} for modelling the expected background noise at the positron detectors. In the simulations, only the electrons that actively contribute to the background noise generation were used. Hence, from the theoretical energy spectrum after the electron-laser interaction as shown in \Fref{fig. electron spectrum and divergence}a), just the $10^7$~primary electrons in the energy range between~1~GeV to 12.8~GeV were included, while the unperturbed~13~GeV electrons were presumed to propagate to the dump and therefore be outside the gating window of the detectors.

\begin{figure}[htp]
    \centering
    \includegraphics[width=0.8\linewidth]{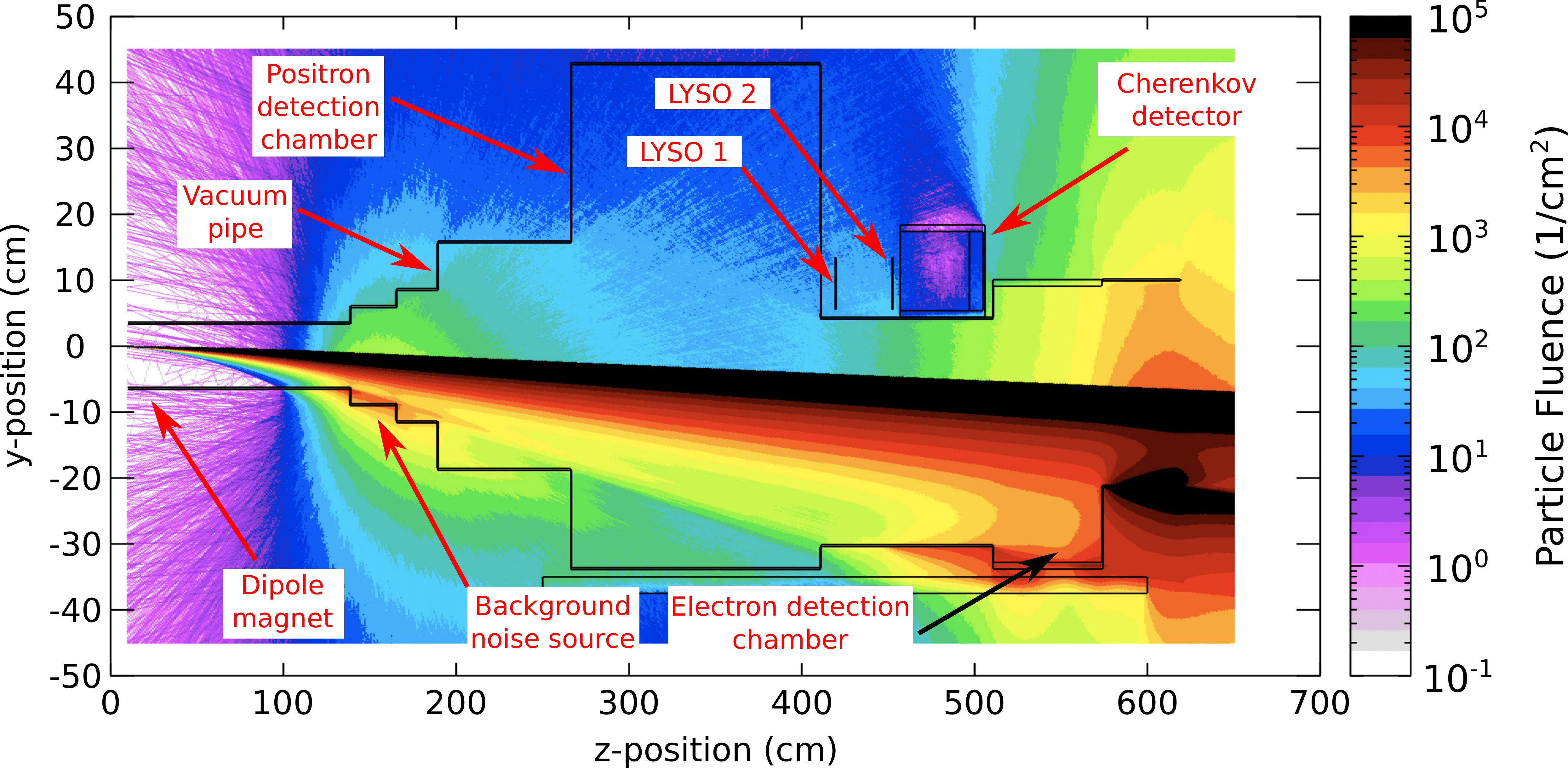}
    \caption{Expected particle fluence (gamma photons, electrons and positrons) at the FACET-II tunnel where the single particle detection system is installed. A shower of secondary particles from the interaction between the deflected primary electron beam with the chamber and vacuum pipe walls travels directly to the detectors generating background noise which cannot be gated.}
    \label{fig. E320 particle fluence map detection region}
\end{figure}

\Fref{fig. E320 particle fluence map detection region} presents the simulated particle fluence at the detection region, corresponding to the layout in~\Fref{fig. FACET detecting region}. The FLUKA simulations include only beamline components within distances where the generated background noise reaches the detectors within the temporal gating window. Consequently, the backscattered noise from the beam dump, which takes longer to reach the detectors, is suppressed on the simulations. Our simulations predict that the majority of background at the detectors arises from the secondary particles generated due to the interaction of low energy scattered electrons with the bottom of the vacuum chamber.

\begin{figure}[htp]
    \centering
    \includegraphics[width=0.45\linewidth]{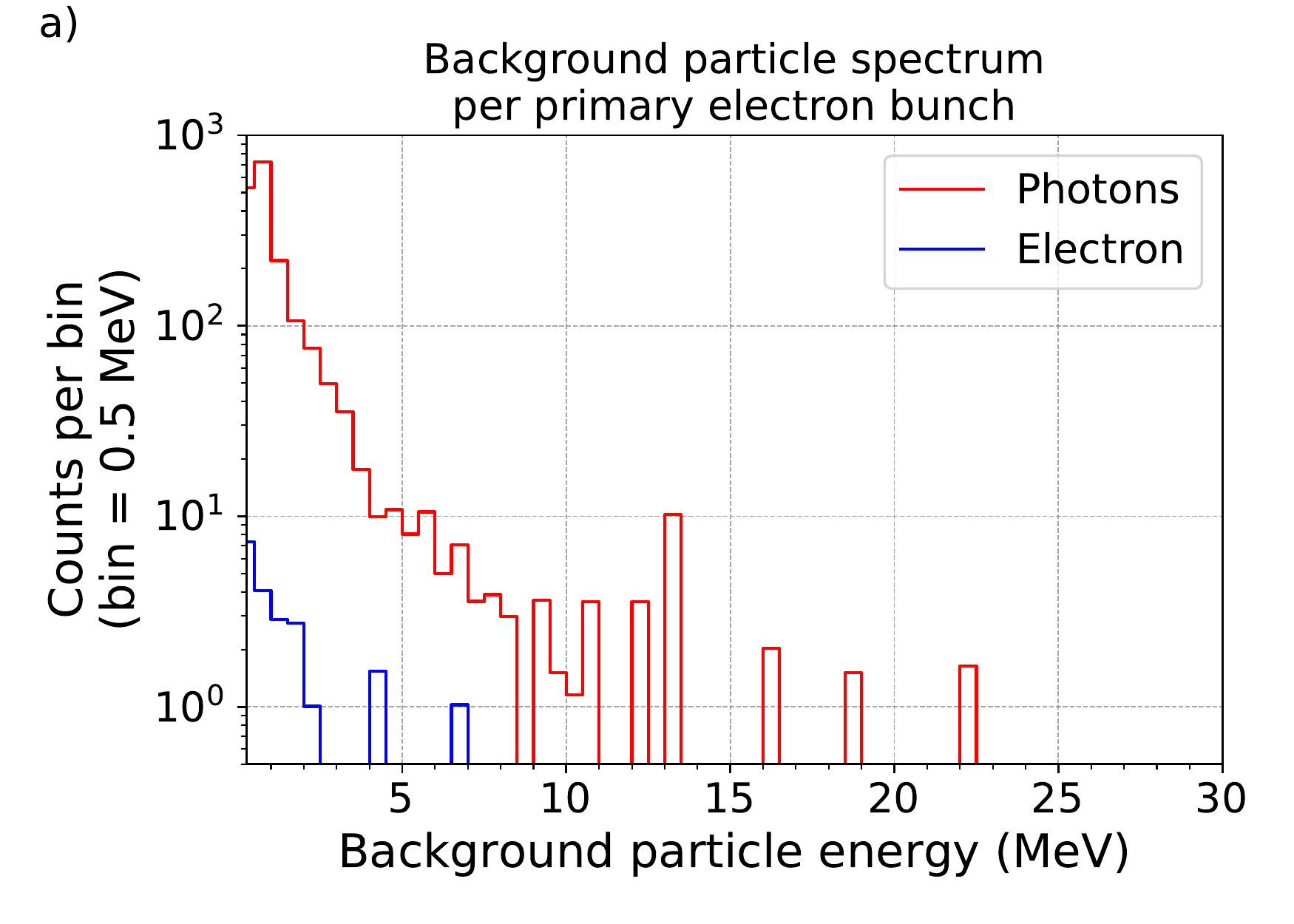}\quad
    \includegraphics[width=0.45\linewidth]{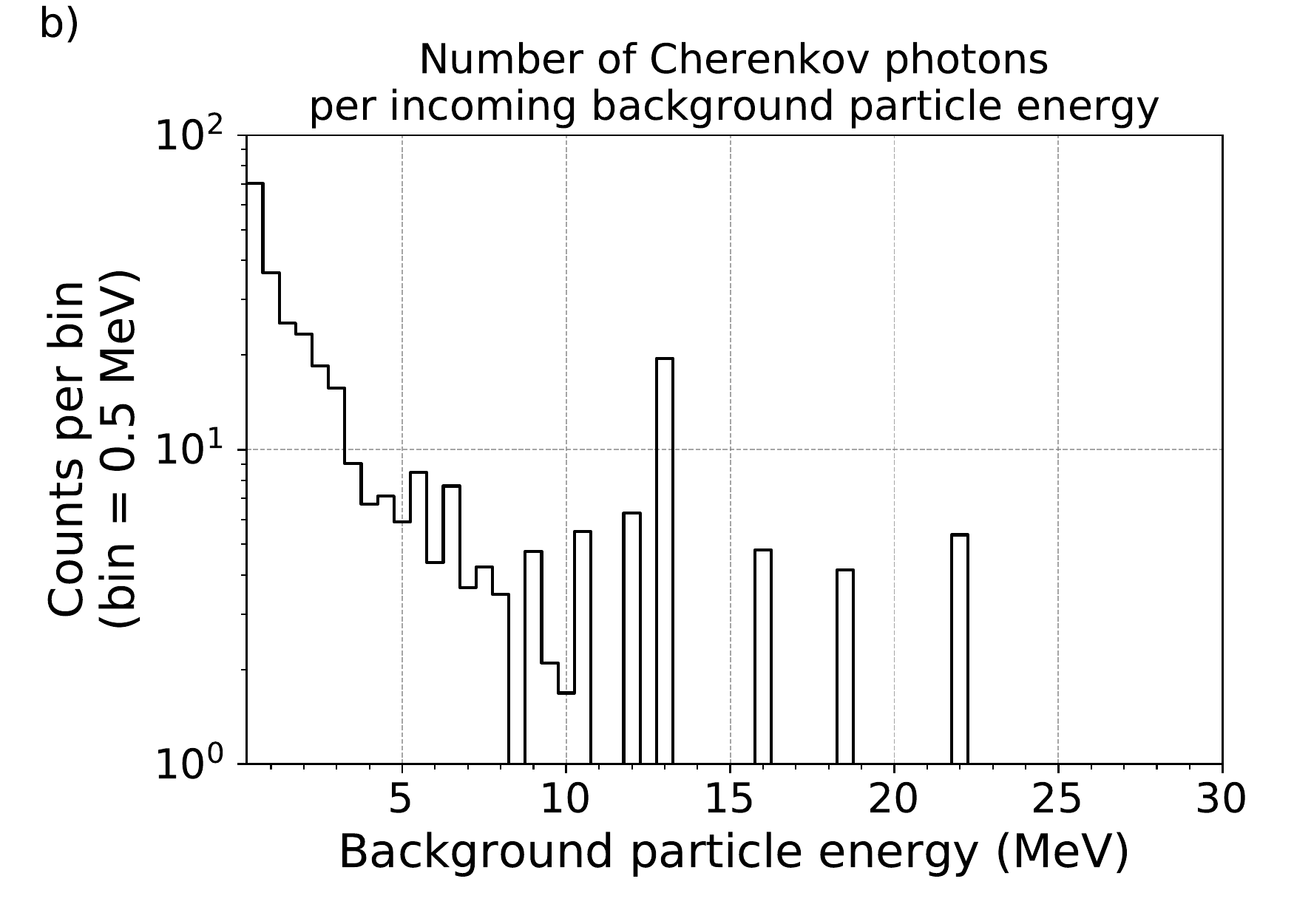}
    \caption{Background spectrum at a single Cherenkov calorimeter detection channel at the Experiment-320: a) Background noise particle spectrum, most of the noise have energies below~25~MeV; b) Number of Cherenkov photons produced per energy of incoming background particle, a total of 320~photons are detected at the single Cherenkov calorimeter channel.}
    \label{fig. Fluka noise particle spectra}
\end{figure}

The particle spectrum of the background incident on the Cherenkov calorimeter is shown in~\Fref{fig. Fluka noise particle spectra}a). As can be seen, the incoming background is mainly composed by photons with energy lower than~25~MeV. The number of Cherenkov photons detected inside a calorimeter channel per incoming particle energy is presented in~\Fref{fig. Fluka noise particle spectra}b). Summing up all Cherenkov photons we find a total of $N_{BG}=\mathrm{320}$ photons per primary electron bunch with a standard variation of $\sigma_{BG}=\mathrm{12}$. From Monte-Carlo simulations, a single~3~GeV positron (characteristic of expected positron energies) is expected to produce only $N_{SIG}=\mathrm{537}$~photons that are detected (see \Sref{sec. calibration of the Cherenkov calorimeter}). 
The expected signal-to-noise ratio (SNR) in terms of detected photons is therefore $\SNR_P = N_{SIG}/N_{BG}=1.7$ i.e. the number of optical photons within the detection channel is approximately tripled when a signal positron is present. The background signal considered in these simulations is produced by the interaction of laser scattered electrons with the vacuum system. It consists of a large number of low energy events generated at some distance from the positron detector and therefore irradiates the Cherenkov detector stack with a smoothly varying flux and low statistical fluctuation. 

The number of detected photons by the PMT $N_{det}$ is given by the sum of the signal produced by a single particle $N_{SIG}$ added to the signal produced by the background noise $N_{BG}$ and the expected uncertainty {\color{\revcolor} $<\sigma_{det}>^2 \,=\, \sigma_{SIG}^2 + \sigma_{BG}^2$} which accounts for the variance of both signal and background,

\begin{equation}
{\color{\revcolor}
    N_{det} = N_{SIG} + N_{BG} \, \pm <\sigma_{det}> \, .
    }
\end{equation}

During the experiment, the background noise $N_{BG}$ is monitored by the calorimeter reference channels at each shot and its value is subtracted from the signal detected $N_{det}$ by the PMTs. Thus, the signal produced by a single particle hit is calculated as {\color{\revcolor}$N_{SIG} = N_{det} - N_{BG} \, \pm <\sigma>$ with uncertainty given by simply propagating the variances such that $<\sigma>^2\,=\, <\sigma_{SIG}>^2 +\, 2<\sigma_{BG}>^2$}. The uncertainty of the signal $\sigma_{SIG}$ is the main contribution on the expected variance $<\sigma>$ for a smoothly varying background consisting of many lower energy particles. Using the calibration given in \Fref{fig. Calibration detected photons at calorimeter channel}, the energy of the particle is evaluated as $E = (N_{SIG}/0.13)^{(1/1.04)}$ and its precision is given by the expected uncertainty $<\sigma>$. The more important SNR measure is therefore the alternative definition of signal to variance $<\sigma>$. This is predicted to be {\color{\revcolor}$\SNR_\sigma=N_{SIG}$/$<\sigma>$ $\approx \mathrm{18}$} for a 3~GeV signal particle, {\color{\revcolor} where the number signal photons of $N_{SIG} = 537$ is obtained from the calorimeter calibration curve for a single 3~GeV particle, and $<\sigma> \approx 29$ is calculated using the the simulated background uncertainty of
$<\sigma_{BG}> = 12$ and signal variance of the number of $N_{SIG}$ given as $<\sigma_{SIG}> = \sqrt{N_{SIG}} \approx 23$ such that $<\sigma> = (2 \cdot 12^2 + 23^2)^{0.5} \approx 29$. The high $\SNR_\sigma >> 1$ demonstrates that individual positrons can clearly be separated from the background.}

\begin{figure}[htp]
    \centering
    \includegraphics[width=0.9\linewidth]{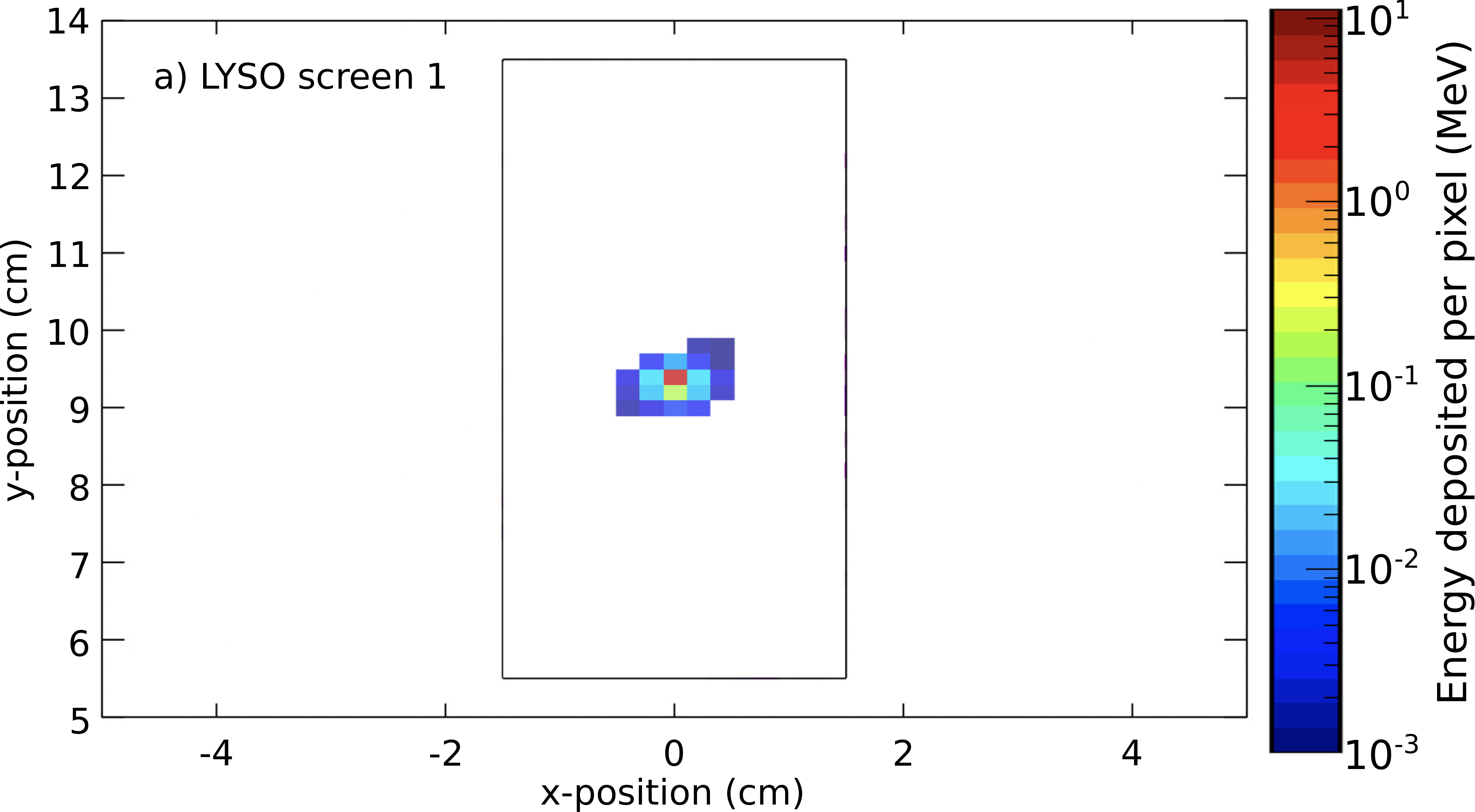}\par\vspace{30pt}
    \includegraphics[width=0.9\linewidth]{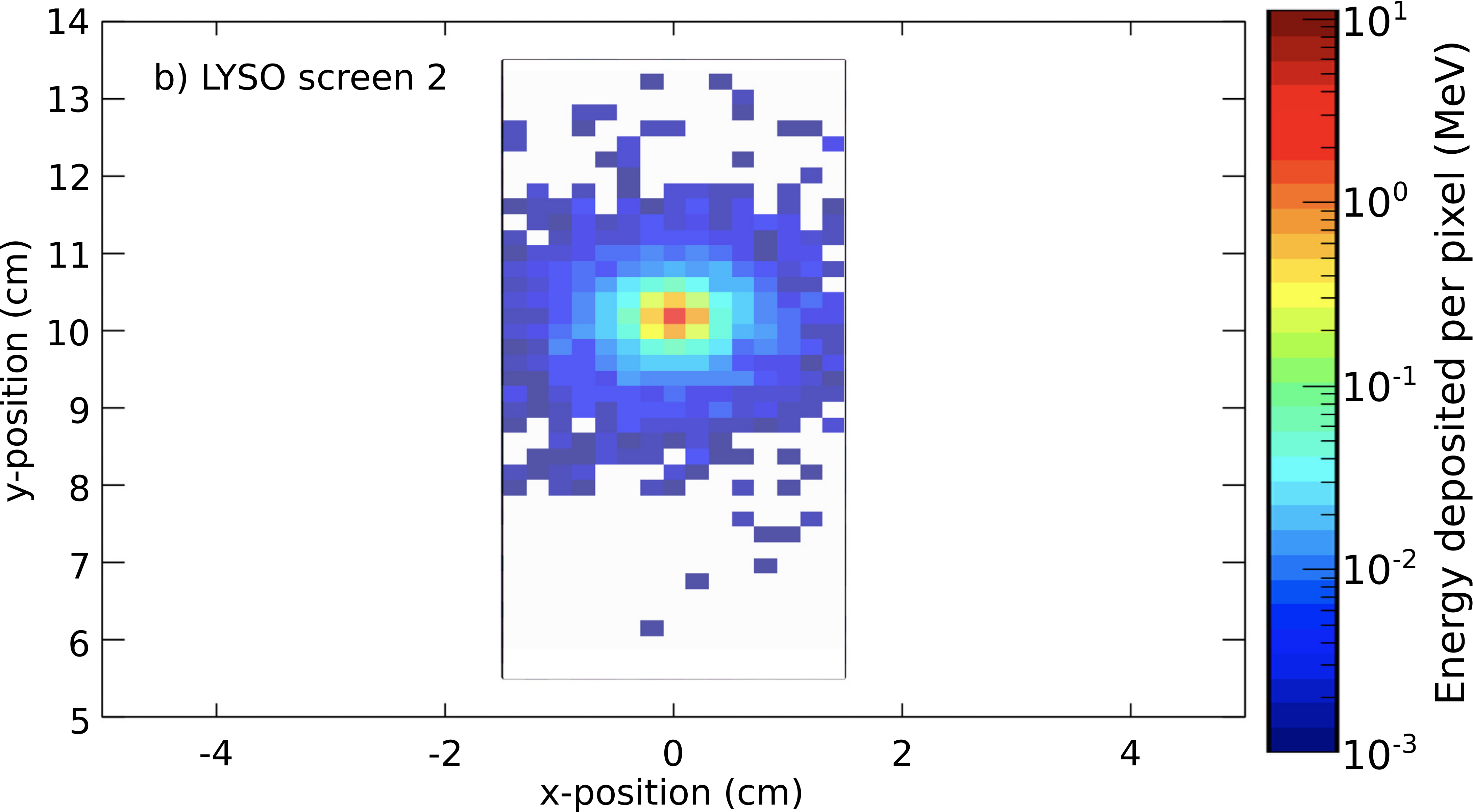}
    \caption{Energy deposited at the LYSO screens at the Experiment-320 after a single particle propagate through them: a) LYSO screen~1; b) LYSO screen number~2. The particle deposits about 5.5~MeV in a single pixel at $\mathrm{y}= 9.4~\mathrm{cm}$ on screen-1 and in a pixel with $\mathrm{y}=10.3~\mathrm{cm}$ at screen-2. The propagation angle of the particle is 28~mrad given the distance between the screens of 330~mm.
    }
    \label{fig. Fluka LYSO screens particle hit}
\end{figure}

The higher spatial resolution of the LYSO crystals enables pair spectra to be measured with a resolution of~60~MeV at 3~GeV for nominal dipole magnet settings.  A simulation of a single 3~GeV particle propagating through the LYSO screens without considering the background radiation is shown in \Fref{fig. Fluka LYSO screens particle hit}. The single particle is deflected by the dipole magnet and propagates  to the LYSO screen-1 depositing about 5.5~MeV. The shower produces signal in a cluster of pixels  on screen-2 with 8.8~MeV deposited.  Recording a high resolution spectrum at low count rates of $< 1$ pair per shot requires integration of many shots and therefore efficient rejection of background events.

\begin{figure}[htp]
    \centering
    \includegraphics[width=0.65\linewidth]{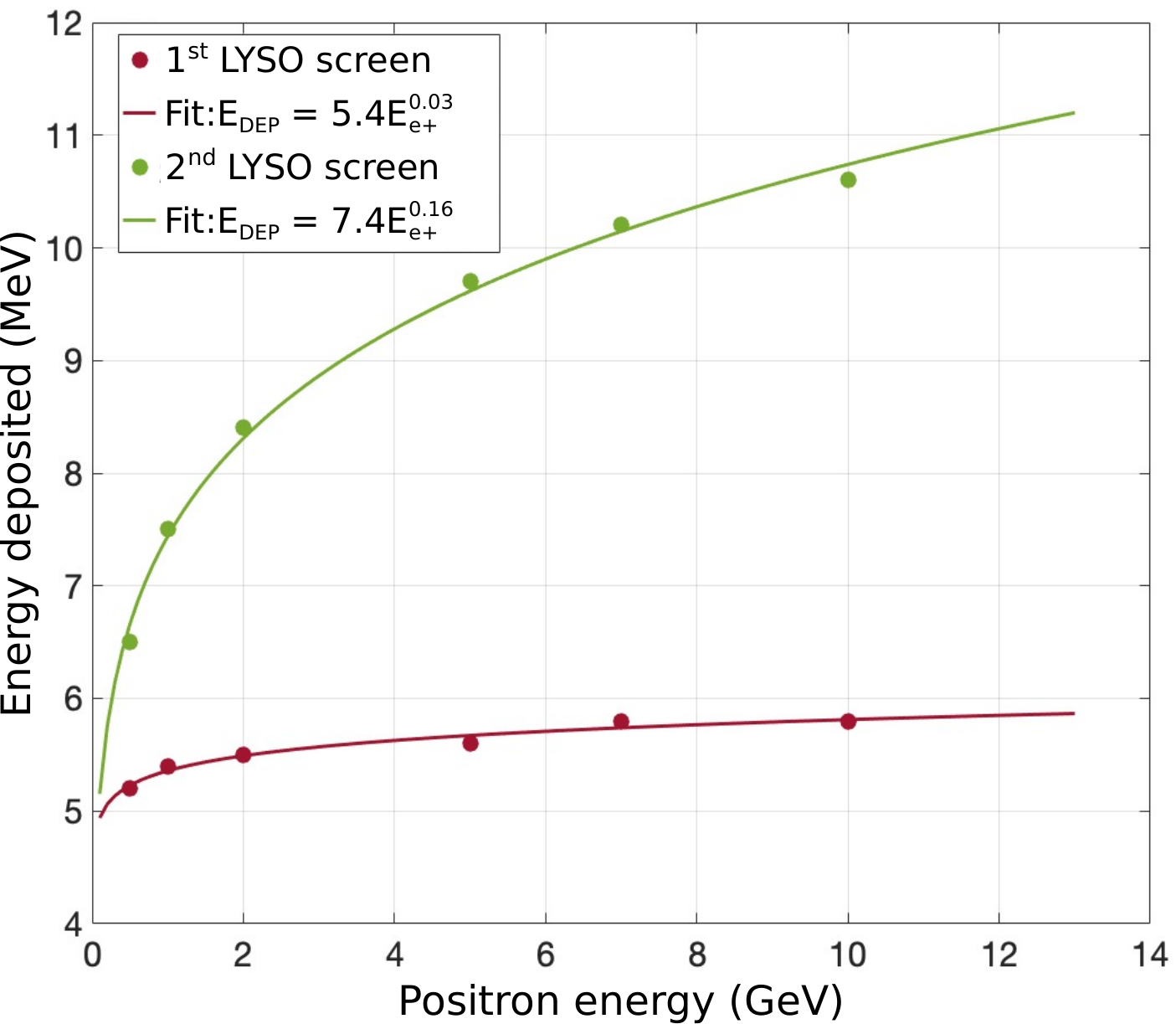}
    \caption{Energy deposited at each LYSO scintillating screen per incoming single particle. As the incoming particle travels through the first crystal screen, secondary particles with lower energy are created increasing the energy deposited at the second screen in comparison with the first LYSO. The typical uncertainty on the data points is on the order of 3-5\%.}
    \label{fig. Fluka Edep at the screens}
\end{figure}

To this end, the energy deposited in each LYSO screen by a single particles hit was evaluated using FLUKA as shown in \Fref{fig. Fluka Edep at the screens}. For 5.5~MeV deposited on a single crystal pixel, about $1.4 \times 10^5$~scintillation photons are produced however only a fraction of them are detected by the camera. The number of detected photons for a GeV-particle passing through the LYSO screens is estimated as $1.4 \times 10^5~\uphotons \cdot \mathrm{CE}_{PCX} \cdot \mathrm{G}_{int} \cdot \mathrm{CE}_{Orca} \cdot \mathrm{QE}_{Orca} \geq 546~\uphotons$, where $\mathrm{CE}_{PCX}$ is the collection efficiency of the PCX condenser lens imaging system calculated as $(\pi \cdot 125^2)/ (800^2) /(4\pi) = 6.1\times 10^{-3}$, $\mathrm{CE}_{Orca}$ is the collection efficiency of the camera $\mathrm{CE}_{Orca} = (\pi \cdot 20^2)/ (250^2) /(4\pi) = 1.6 \times 10^{-3}$,  $\mathrm{QE}_{Orcas} \approx 0.4$ is the quantum efficiency of the ORCA-Flash camera at the scintillation light wavelength of 410~nm~\cite{Orcas.2021}, and the parameter $\mathrm{G}_{int} = 10^3$ is the gain applied on the dual microchannel plate (MCP) of the image intensifier already taken into account the quantum efficiency of the device. Following the same calculation method, we expect an average of 858~scintillation photons being detected for the LYSO screen-2. {\color{\revcolor} The uncertainty on the LYSO measurements is determined by the overall counting efficiency (collection and quantum) which is to be calibrated after FACET-II be commissioned.}

To reject the background we require a valid event to follow the calculated particle track to within one pixel  and  define the threshold to be within $3 \sigma$ of the expected signal level in each of the three detectors (screen-1, screen-2 and the Cherenkov calorimeter), leading to $>99 \%$ of true events being counted and therefore $<1 \%$ false negatives. 

\begin{figure}[htp]
    \centering
    \includegraphics[width=0.9\linewidth]{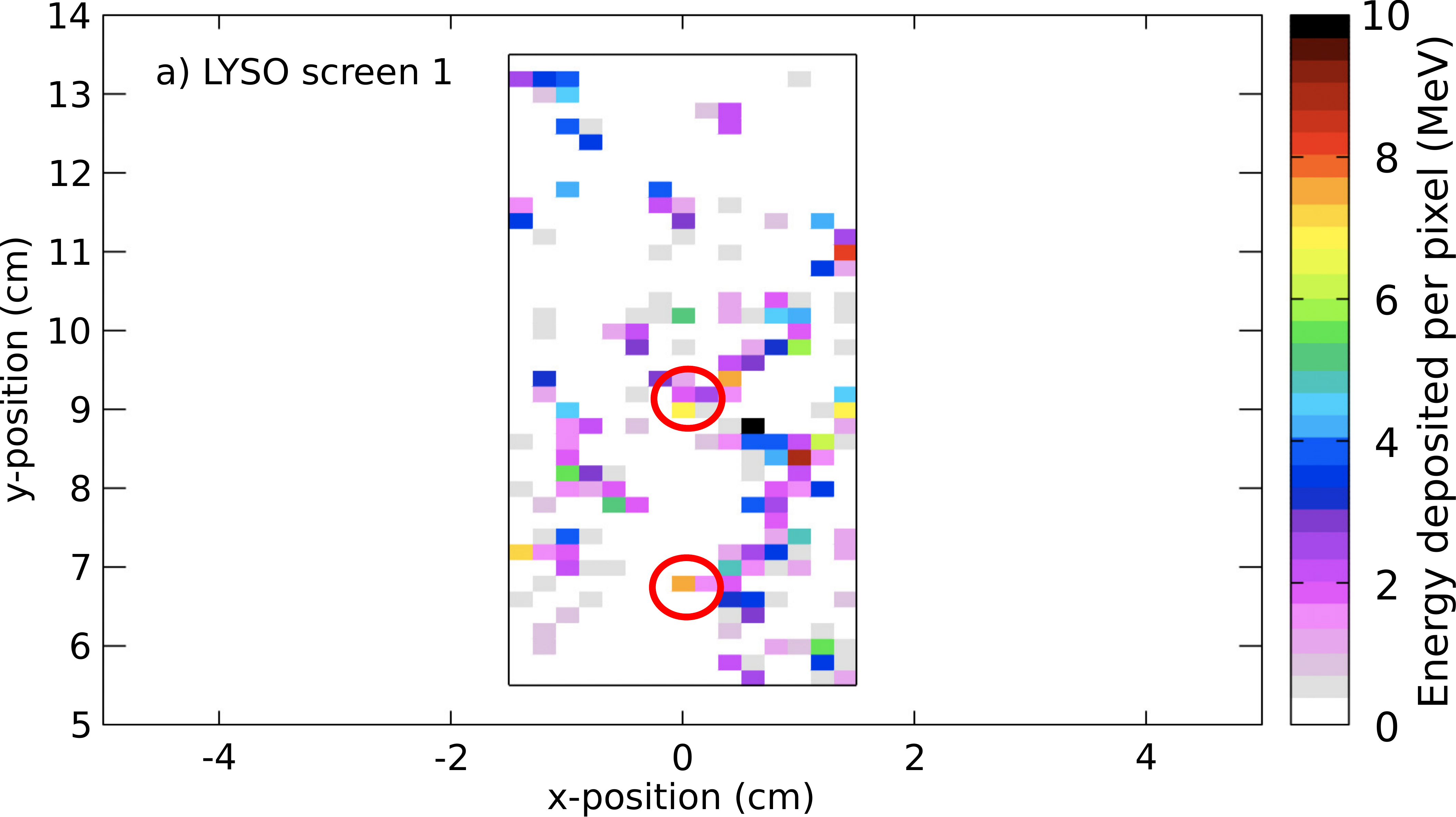}\par\vspace{30pt}
    \includegraphics[width=0.9\linewidth]{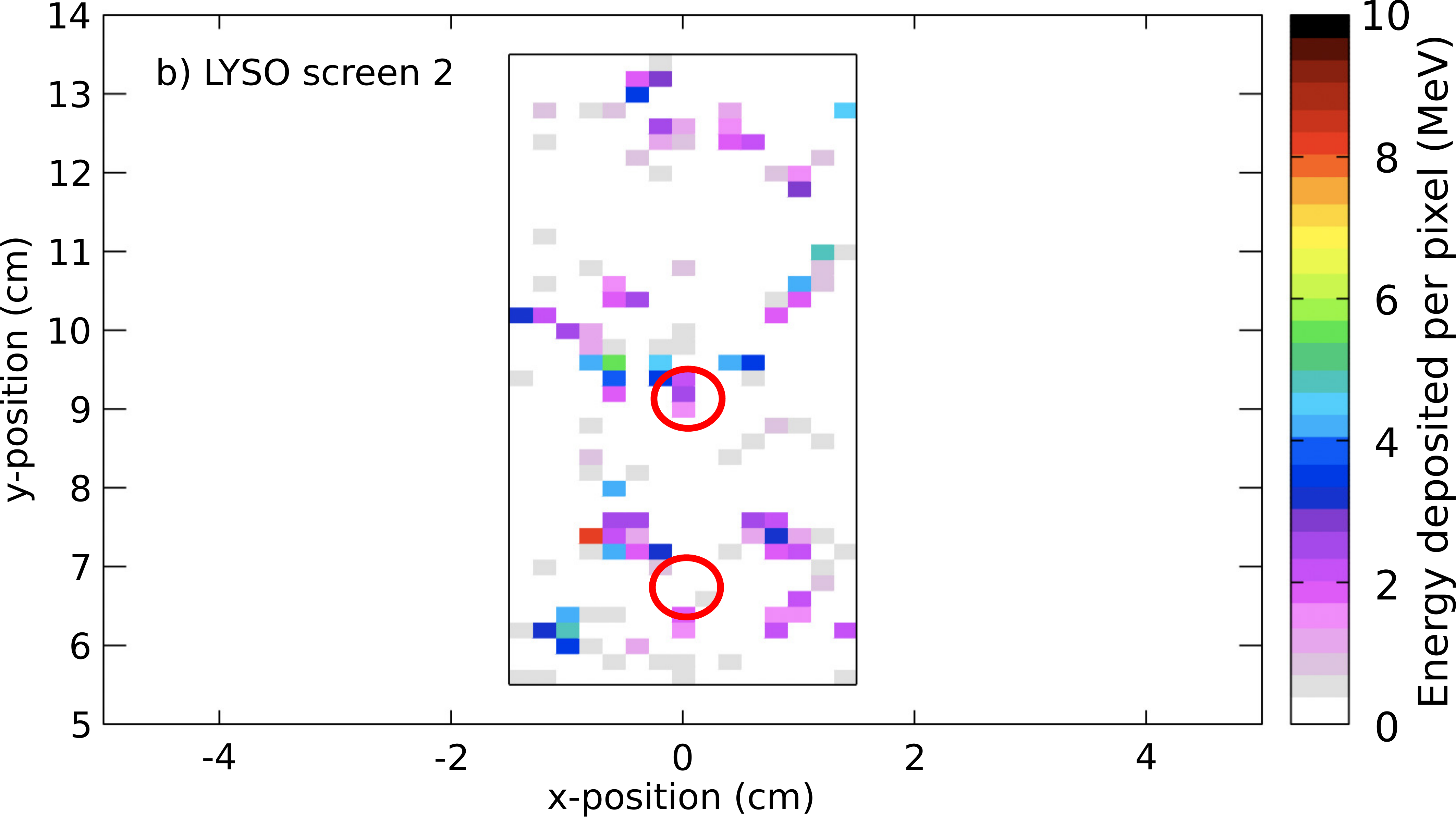}
    \caption{Energy deposited at the LYSO screens at the Experiment-320: a) LYSO screen~1; b) LYSO screen number~2. All background events (with exception of event highlighted by a red circle) on screen-1 can be rejected due to energy deposited per pixel on screen-1 or absence of $> 6~\mathrm{MeV}$ pixel cluster on screen-2. The highlighted pixel in the centre of screen-1 and area on screen-2 would be consistent with a $> 1~\mathrm{GeV}$ incident positron, but can be rejected due to the calorimeter signal.
    }
    \label{fig. Fluka LYSO screens}
\end{figure}

Simulations to evaluate the background noise level on the screens were also performed. Most of the background hits shown in \Fref{fig. Fluka LYSO screens} on screen-1 and screen-2 can be rejected based on the above energy thresholds alone. The highlighted pixels on screen-1 with energy deposited $> 6~\mathrm{MeV}$ can be neglected by analysing their tracked pixels on screen-2. The absence of cluster of pixels with integrated energy deposited $> 8~\mathrm{MeV}$ as highlighted in \Fref{fig. Fluka LYSO screens}b) strongly indicates that the hits on screen-1 are originated from background noise. Finally, none of the background events shown meet the detector threshold in the Cherenkov detector.

These simulations only account for background generated by the electrons scattered due to the electron beam-laser interaction. {\color{\revcolor} The measurement of false positive events is only possible if the calorimeter measures a sufficiently high signal above the background. However, our FLUKA simulations show that the background noise consists of a bath of many low energy events. Therefore, the reference channels and signal channels prevent false positives arising from the low energy particle background. Thus, based on our simulation inputs, the false positive rate would be zero. To estimate the real false positive rate, we need to be able to evaluate the probability of giving a false positive on the Cherenkov calorimeter by having a high-energy localized event on the detector (or split event). To quantify this probability, the background radiation level of hard gamma photons in FACET-II needs to be known, which is will be possible as soon as FACET-II becomes operational.
}


\section{Conclusions}
In this paper, a single particle detection system designed to measure positron spectra for strong-field QED experiments is presented. The implementation in the upcoming Experiment-320 at FACET-II (SLAC) is detailed with calibration data from the ELBE accelerator. Based on this calibration a single~3~GeV particle hit at the Cherenkov detector generates about~537 detectable photons by the calorimeter with an energy resolution of $\approx20~\%$. Furthermore, Monte-Carlo simulations were performed to demonstrate that a signal-to-noise ratio of {\color{\revcolor}$\SNR_\sigma = 18$} is predicted for the Cherenkov calorimeter detector at the Experiment-320. The combination of the LYSO screens with the Cherenkov detector allows efficient rejection of background events and the recording of positron spectra with $\Delta E/E=0.02$ even for pair production rates of $\ll 1$ per shot. 

\ack
This work has been funded by the Deutsche Forschungsgemeinschaft (DFG) under Grant No. 416708866 within the Research Unit FOR2783/1. Parts of this research were carried out at ELBE at the Helmholtz-Zentrum Dresden - Rossendorf e. V., a member of the Helmholtz Association. We would like to thank the ELBE and FACET staff for their assistance. This work was supported by the U.S.\ Department of Energy under contract number DE-AC02-76SF00515. PHB, EG, EI, and DAR were supported by the U.S.\ Department of Energy, Office of Science, Office of Fusion Energy Sciences under award DE-SC0020076. GS would like to acknowledge support from the EPSRC (grant numbers: EP/P010059/1, EP/T021659/1, and EP/V044397/1).

\appendix
\section{LYSO:Ce decay time calibration}
\label{sec: LYSO:Ce decay time calibration}

\begin{figure}[htp]
    \centering
    \includegraphics[width=0.45\linewidth]{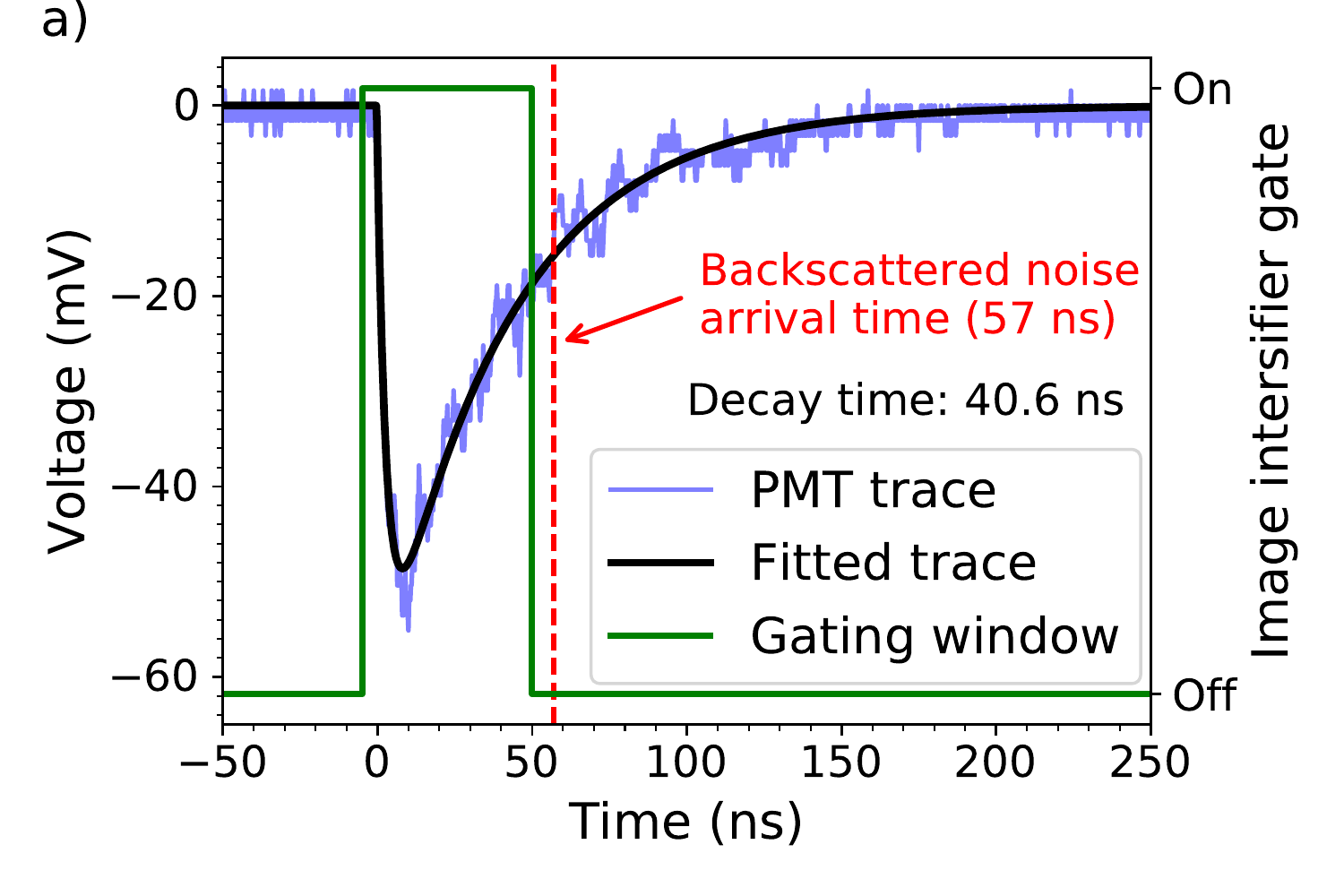}
    \includegraphics[width=0.45\linewidth]{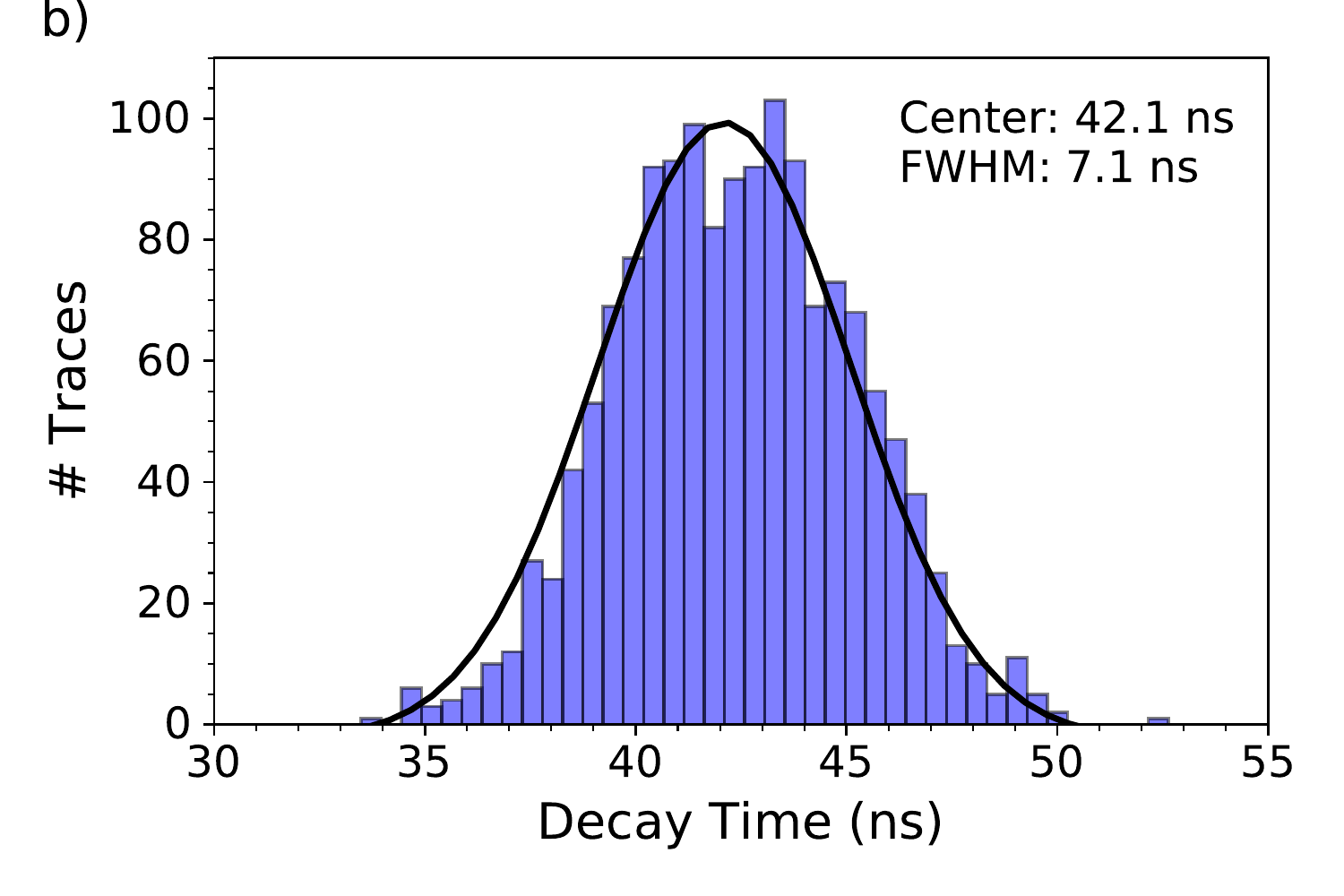}
    \caption{Calibration of the decay time of the LYSO screens. (a) Shows a single scintillation trace used to evaluate the decay time of the scintillation signal, the image intensifier gate window is also shown demonstrating that the LYSO scintillation light can be captured without the background noise influence. (b) Presents the histogram distribution of the decay time for 1500 events where an average decay time of 42.2~ns with FWHM of 7.1~ns is calculated.}
    \label{fig. LYSO decay time}
\end{figure}

The decay time of the LYSO screens were calibrated using a radioactive Sodium-22 source to produce scintillation photons in the LYSO screens and a photomuliplier tube to diagnose the scintillation photons. A total of 1500 traces captured by the PMT were analysed and a curve of the form 

\begin{equation}
    V = A \left[ \exp\left( -\frac{(t - t_0)}{\tau}\right) - \exp\left( -\frac{(t - t_0)}{\tau_p}\right)\right] \Theta(t - t_0)
    \label{eq. PMT trace fit}
\end{equation}
was fitted on each captured trace. In \Eref{eq. PMT trace fit}, $A$ is the signal amplitude, $\tau$ is the decay time of the signal, $\tau_P$ stands for the signal rise time and $t_0$ is time shift of the signal. The function $\Theta(t)$ is the step-function which allows to shift the signal in time. An example of a single captured trace used on the evaluation of the decay times is shown in \Fref{fig. LYSO decay time}a).

\Fref{fig. LYSO decay time}b) shows a histogram distribution of the evaluated decay time $\tau$ of each captured signal. A normal distribution was fitted on the histogram data and an average decay time of the LYSO crystals of 42.2~ns with FWHM of 7.1~ns was calculated. Both results are in agreement with measurements reported in the literature~\cite{Seifert.2012}. 

In \Fref{fig. LYSO decay time}a), the image intensifier temporal gate window is also shown. The capturing window starts shortly before the single particle arrives and ends few-ns before the arrival of background noise from the beam dump allowing to capture most of the scintillation light from the LYSO screens without overlapping with the background noise.

\section{Dark Current Calibration}
\label{sec: Dark Current Calibration}
To estimate the number of electrons at each accelerator dark current shot, Gafchromic EBT3 Radiochromic films were used~\cite{EBT3.2020}. The RCF films were placed in front of the Cherenkov calorimeter detector and the absorbed dose in the films was recorded for two define time periods,~3270~s and~650~s.

The RCFs were digitized using a commercial scanner model Epson Perfection~V750 Pro~\cite{Epson.2020}. The chosen flatbed scanner allows to store information of 16-bit RGB color information from the scanned film. By calibrating the scanner transmission light, one can retrieve the optical density of the irradiated RCFs by using the digitized color counts for each individual RGB channel. From the optical density, information retrieved from the RCFs, the absorbed dose is obtained~\cite{wulff_construction_2020}. The total number of electrons that have passed through a selected area of an RCF is given by~\cite{wulff_construction_2020}
\begin{equation}
    N_e = \frac{d\,\rho}{e \, E_{abs}} A_{p} \sum_{i,j}\left[ D(C_{R})_{i,j} - D_{bkg}\right] \, ,
    \label{eq. Ne RCF}
\end{equation}
where $D_{bkg}$ is the average background dose noise on the RCF, $e$ is the electron charge, $d = 28$~\microm~and $\rho = 1.20~\mathrm{g/cm}^3$ are the thickness and density of the active layer of the radiochromic film, $A_{p} = \left( 25.4~\mathrm{mm} / 300\right)^2$ is the area for one pixel, and, $D(C_{R})_{i,j}$ is the total absorbed dose at the pixel given by the indexes $(i, j)$.

\begin{figure}[htb]
    \centering
    \includegraphics[width=0.45\linewidth]{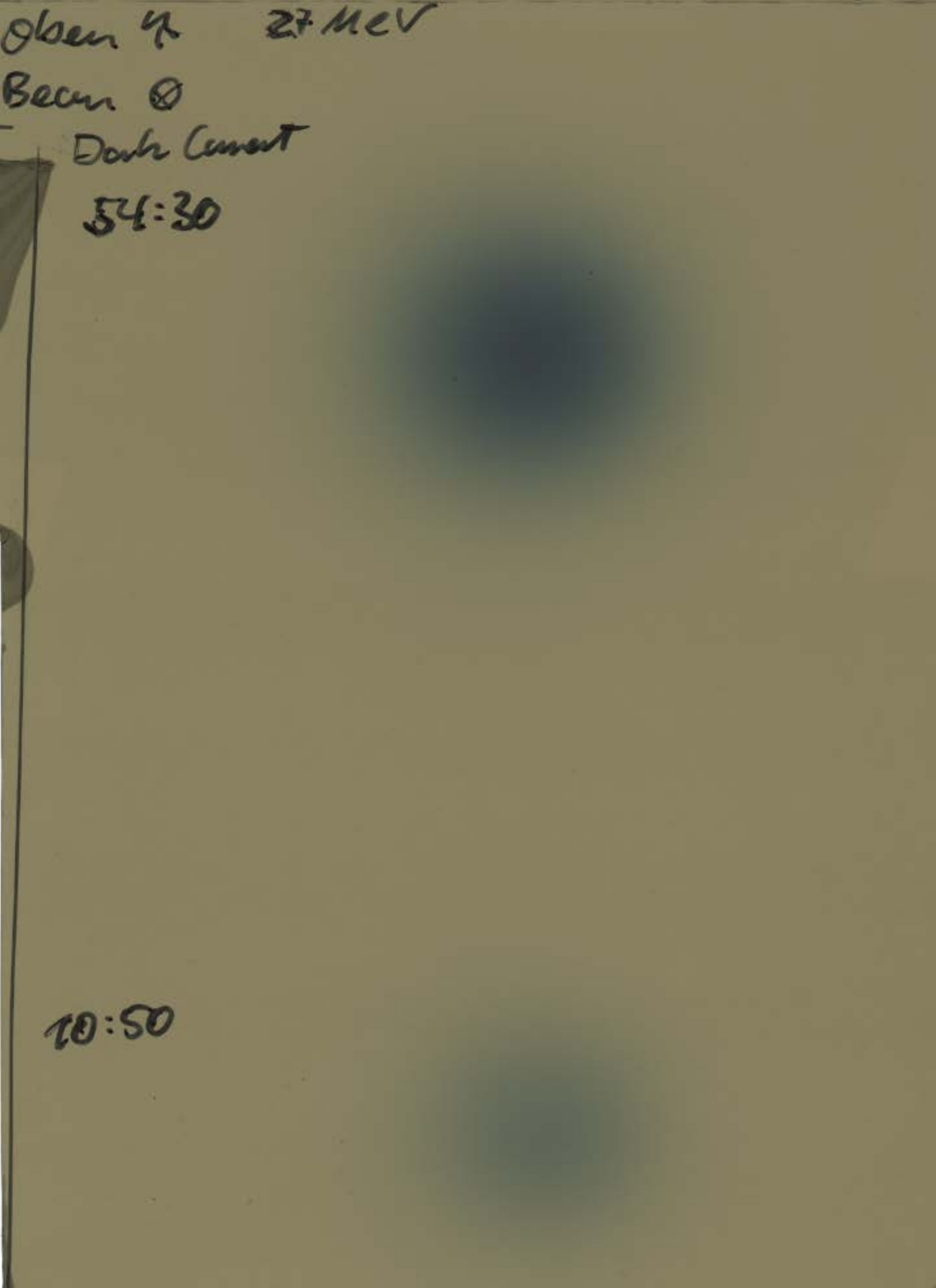}
    \caption{Radiochromic film used for determining the dark current of the ELBE accelerator.}
    \label{Fig. RFC dark current}
\end{figure}

The dark current of the accelerator is considered as a constant background source or noise on the measurements and has a frequency of 260~MHz. At the charge calibration measurements, the noise given by the dark current needs to be taken into account. The dark current can be measured as soon as the radio frequency (RF) injector gun was activated by opening the accelerator shutter. The total number of dark current electrons can be calculated using the following equation,
\begin{equation}
    N_e^{dc} = N_e^c \times 260~\mathrm{MHz} \times T \, ,
\end{equation}
where $T$ is the total RCF exposure time to the dark current, $N_e^c$ is the average number of electrons in one dark current cycle. The RCF used for the dark current measurement is shown in Fig.~\ref{Fig. RFC dark current}.

By scanning the RCF shown in Fig.~\ref{Fig. RFC dark current} with the Epson Perfection~V750 Pro flatbed scanner and post processing the individual color channels, the two available measurements of the dark current were analyzed. The first measurement was irradiated for a period of time of approximately 11~min and resulted on an average of 0.164 electrons per 3.846~ns, equivalently to a current of $I_{dc}^{11} = 6.81~\mathrm{pA}$. The second dark current measurement spot was irradiated for a period of 54.5~min and yield an average of 0.150~electrons per 3.846~ns, or $I_{dc}^{54.5} = 6.23~\mathrm{pA}$. 

\subsection{Uncertainty Evaluation of the Dark Current}
In calibration experiment using the dark current of the ELBE accelerator, we identify the following possible sources of uncertainties: 

\begin{itemize}
    \item Background noise on the measurement
    \item Large energy spread of the electron beam
    \item Uncertainty on the digitalization of the colour channels of the Radiochromic films (RCF) by the flatbed scanner
    \item Uncertainty on the conversion between the measured optical density from the RCFs to absorbed dose
\end{itemize}

The background noise on the measurement can be neglected since there cannot be any other source of background scattered particle inside the experimental cave at ELBE and the signals captured by the PMTs were triggered above the typical background levels of the photomultiplier tubes as well as its dark current. The large energy spread of the electron source can also be neglected since the ELBE accelerator provides single particle energies filtered by a magnetic system and resulting in a energy spread of $< 50~\mathrm{keV}$. Hence, the electron beam has an energy resolution of $E = 50~\mathrm{keV}/27~\mathrm{MeV} < 0.2\%$.

The uncertainty on the digitalization of the colour channels of the flatbed scanner was also addressed during the evaluation of the accelerator dark current. A calibration of the three colour channels (red, green, and blue) of the Epson Perfection V750 flatbed scanner was performed using commercial Kodak Wratten~2 Neutral Density No.~96 filters~\cite{WrattenFilter} which were also calibrated and a difference of $<0.5\%$ from their nominal value was found. In the readout of the radiochromatic films (RCFs), we focused on the red channel of the scanned films and a calibration curve obtained for the red colour channel of the scanner was obtained through a curve fit of the form $\mathrm{OD} = 2.229\, \exp(-C_R/21957) - 0.427$ where OD is the optical density and $C_R$ is the digitized intensity in counts of the red colour channel by the flatbed scanner. The fitted curve presented a $R^2=0.9996$ which confirms that the fitted curve agrees well the calibration data points. 

The conversion between optical density of the RCF obtained at the red colour channel read by the scanner and the absorbed dose was also evaluated. RCFs films were illuminated by a proton source (Jena University Laboratory for Ion Acceleration tandem accelerator, namely JULIA) with a known fluence which provides a specific absorbed dose on the films. Later, the illuminated RCFs were scanned and a conversion between the optical density at the red colour channel and the absorbed dose was be evaluated~\cite{wulff_construction_2020, MAX_EBT_Calibration}. As result, a transfer function between the digitized intensity $C_R$ and the absorbed dose by the RCF was obtained as,

\begin{equation}
\mathrm{Dose (Gy)} = \frac{6.1 \, C_R - 0.221\times 10^6}{5.3\times 10^3 - C_R} \, .
\end{equation}

The transfer function above has a fluence error of 5\% for the dose applied from the proton source and the other uncertainties are assumed to be negligible. As only the fluence error of 5\% is the only substantial uncertainty, the value of 5\% is also applied for the number of electrons from the dark current.

Finally, the ELBE dark current measurement of 650~seconds provides a current value of $(0.164 \pm 0.008)$ electrons per RF cycle and the longer measurement of 3270~seconds results in a dark current of $(0.150 \pm 0.007)$ electrons per RF cycle.

{\color{\revcolor}From both measurements, the best estimate for the dark current value is calculated using the weighted average method as described in~\cite{Taylor.1997},
\begin{equation}
    x_{wav} = \frac{\sum_i \, w_i x_i}{\sum_i \, w_i},
\end{equation}
where $x_i$ is the i-th measured value with uncertainty of $\sigma_i$, and $w_i$ is the weight calculated as $w_i = 1/\sigma_i^2$. The uncertainty $\sigma_{wav}$ in $x_{wav}$ is calculated using error propagation such that
\begin{equation}
    \sigma_{wav} = \frac{1}{\sqrt{\sum_i\, w_i}} \,.
\end{equation}

Hence, the best estimate for the dark current is $x_{wav}^{DC} = (x_{wav} \pm \sigma_{wav}) = (0.156 \pm 0.005)$ electrons per RF cycle.
}

\section*{References}
\bibliography{iopart-num}

\providecommand{\newblock}{}
\begin{thebibliography}{10}
\expandafter\ifx\csname url\endcsname\relax
  \def\url#1{{\tt #1}}\fi
\expandafter\ifx\csname urlprefix\endcsname\relax\def\urlprefix{URL }\fi
\providecommand{\eprint}[2][]{\url{#2}}

\bibitem{karshenboim2006study}
Karshenboim S~G, Eidelman S, Fendel P, Ivanov V, Kolachevsky N, Shelyuto V and
  H{\"a}nsch T 2006 {\em Nucl. Phys. B, Proc. Suppl.\/} {\bf 162} 260--263

\bibitem{Ritus.1985}
Ritus V~I 1985 {\em J. Sov. Laser Res.\/} {\bf 6} 497--617

\bibitem{diPiazza2012}
Di~Piazza A, M\"uller C, Hatsagortsyan K~Z and Keitel C~H 2012 {\em Rev. Mod.
  Phys.\/} {\bf 84}(3) 1177--1228

\bibitem{Aaboud2017}
Aaboud M~e~a 2017 {\em Nat. Phys.\/} {\bf 13} 852--858 ISSN 1745-2473

\bibitem{Nielsen2020}
Nielsen C~F, Justesen J~B, S\o{}rensen A~H, Uggerh\o{}j U~I and Holtzapple R
  (CERN NA63 Collaboration) 2020 {\em Phys. Rev. D\/} {\bf 102}(5) 052004

\bibitem{Yakimenko2019}
Yakimenko V, Meuren S, Del~Gaudio F, Baumann C, Fedotov A, Fiuza F, Grismayer
  T, Hogan M~J, Pukhov A, Silva L~O and White G 2019 {\em Phys. Rev. Lett.\/}
  {\bf 122}(19) 190404

\bibitem{strickland1985compression}
Strickland D and Mourou G 1985 {\em Opt. Commun.\/} {\bf 55} 447--449

\bibitem{Mourou2006}
Mourou G~A, Tajima T and Bulanov S~V 2006 {\em Rev. Mod. Phys.\/} {\bf 78}(2)
  309--371

\bibitem{lureau2020high}
Lureau F, Matras G, Chalus O, Derycke C, Morbieu T, Radier C, Casagrande O,
  Laux S, Ricaud S, Rey G {\em et~al.\/} 2020 {\em High Power Laser Sci.
  Eng.\/} {\bf 8}

\bibitem{Gies2018}
Gies H, Karbstein F, Kohlf{\"{u}}rst C and Seegert N 2018 {\em Phys. Rev. D\/}
  {\bf 97} 076002

\bibitem{Heinzl2006}
Heinzl T, Liesfeld B, Amthor K~U, Schwoerer H, Sauerbrey R and Wipf A 2006 {\em
  Opt. Commun.\/} {\bf 267} 318--321 ISSN 00304018

\bibitem{Karbstein.2015}
Karbstein F, Gies H, Reuter M and Zepf M 2015 {\em Phys. Rev. D\/} {\bf 92}
  ISSN 1550-7998

\bibitem{Bula.1996}
Bula C, McDonald K~T, Prebys E~J, Bamber C, Boege S, Kotseroglou T, Melissinos
  A~C, Meyerhofer D~D, Ragg W, Burke D~L, Field R~C, Horton-Smith G, Odian A~C,
  Spencer J~E, Walz D, Berridge S~C, Bugg W~M, Shmakov K and Weidemann A~W 1996
  {\em Phys. Rev. Lett.\/} {\bf 76}(17) 3116--3119

\bibitem{Burke.1997}
Burke D~L, Field R~C, Horton-Smith G, Spencer J~E, Walz D, Berridge S~C, Bugg
  W~M, Shmakov K, Weidemann A~W, Bula C, McDonald K~T, Prebys E~J, Bamber C,
  Boege S~J, Koffas T, Kotseroglou T, Melissinos A~C, Meyerhofer D~D, Reis D~A
  and Ragg W 1997 {\em Phys. Rev. Lett.\/} {\bf 79}(9) 1626--1629

\bibitem{Bamber.1999}
Bamber C, Boege S~J, Koffas T, Kotseroglou T, Melissinos A~C, Meyerhofer D~D,
  Reis D~A, Ragg W, Bula C, McDonald K~T, Prebys E~J, Burke D~L, Field R~C,
  Horton-Smith G, Spencer J~E, Walz D, Berridge S~C, Bugg W~M, Shmakov K and
  Weidemann A~W 1999 {\em Phys. Rev. D\/} {\bf 60}(9) 092004

\bibitem{Reiss.1971}
Reiss H~R 1971 {\em Phys. Rev. Lett.\/} {\bf 26} 1072--1075 ISSN 0031-9007

\bibitem{RALProposal.2010}
Keitel C~H, {Di Piazza} A, Paulus G~G, Stoehlker T, Clark E~L, Mangles S,
  Najmudin Z, Krushelnick K, Schreiber J, Borghesi M, Dromey B, Geissler M,
  Riley D, Sarri G and Zepf M 2010  (arXiv:2103.06059v1)

\bibitem{Cole.2018}
Cole J~M, Behm K~T, Gerstmayr E, Blackburn T~G, Wood J~C, Baird C~D, Duff M~J,
  Harvey C, Ilderton A, Joglekar A~S, Krushelnick K, Kuschel S, Marklund M,
  McKenna P, Murphy C~D, Poder K, Ridgers C~P, Samarin G~M, Sarri G, Symes D~R,
  Thomas A~G~R, Warwick J, Zepf M, Najmudin Z and Mangles S~P~D 2018 {\em Phys.
  Rev. X\/} {\bf 8} 011020

\bibitem{Poder.2018}
Poder K, Tamburini M, Sarri G, Di~Piazza A, Kuschel S, Baird C~D, Behm K,
  Bohlen S, Cole J~M, Corvan D~J, Duff M, Gerstmayr E, Keitel C~H, Krushelnick
  K, Mangles S~P~D, McKenna P, Murphy C~D, Najmudin Z, Ridgers C~P, Samarin
  G~M, Symes D~R, Thomas A~G~R, Warwick J and Zepf M 2018 {\em Phys. Rev. X\/}
  {\bf 8} 031004

\bibitem{meuren2019probing}
Meuren S 2019 Probing strong-field {QED} at {FACET-II} ({SLAC} {E-320})
  \url{https://conf.slac.stanford.edu/facet-2-2019/sites/facet-2-2019.conf.slac.stanford.edu/files/basic-page-docs/sfqed_2019.pdf}

\bibitem{meuren2020probing}
Meuren S 2020 {E-320}: {Probing} {Strong}-field {QED} at {FACET-II}
  \url{https://facet.slac.stanford.edu/sites/facet.slac.stanford.edu/files/E320_PAC2020_Meuren.pdf}

\bibitem{WorkshopSFQED.2019}
Altarelli M, Assmann R, Burkart F, Heinemann B, Heinzl T, Koffas T, Maier A~R,
  Reis D, Ringwald A and Wing M 2019  (arXiv:1905.00059v1)

\bibitem{Hartin.2019}
Hartin A, Ringwald A and Tapia N 2019 {\em Phys. Rev. D\/} {\bf 99}(3) 036008

\bibitem{luxe_letter.2019}
Abramowicz H, Altarelli M, Aßmann R, Behnke T, Benhammou Y, Borysov O,
  Borysova M, Brinkmann R, Burkart F, Büßer K, Davidi O, Decking W, Elkina N,
  Harsh H, Hartin A, Hartl I, Heinemann B, Heinzl T, TalHod N, Hoffmann M,
  Ilderton A, King B, Levy A, List J, Maier A~R, Negodin E, Perez G, Pomerantz
  I, Ringwald A, Rödel C, Saimpert M, Salgado F, Sarri G, Savoray I, Teter T,
  Wing M and Zepf M 2019  (arXiv:1909.00860)

\bibitem{LUXECDR.2021}
Abramowicz H, Acosta U~H, Altarelli M, Assmann R, Bai Z, Behnke T, Benhammou Y,
  Blackburn T, Boogert S, Borysov O, Borysova M, Brinkmann R, Bruschi M,
  Burkart F, Büßer K, Cavanagh N, Davidi O, Decking W, Dosselli U, Elkina N,
  Fedotov A, Firlej M, Fiutowski T, Fleck K, Gostkin M, Grojean C, Hallford
  J~A, Harsh H, Hartin A, Heinemann B, Heinzl T, Helary L, Hoffmann M, Huang S,
  Huang X, Idzik M, Ilderton A, Jacobs R~M, Kaempfer B, King B, Lakhno H,
  Levanon A, Levy A, Levy I, List J, Lohmann W, Ma T, Macleod A~J, Malka V,
  Meloni F, Mironov A, Morandin M, Moron J, Negodin E, Perez G, Pomerantz I,
  Poeschl R, Prasad R, Quere F, Ringwald A, Roedel C, Rykovanov S, Salgado F,
  Santra A, Sarri G, Saevert A, Sbrizzi A, Schmitt S, Schramm U, Schuwalow S,
  Seipt D, Shaimerdenova L, Shchedrolosiev M, Skakunov M, Soreq Y, Streeter M,
  Swientek K, Hod N~T, Tang S, Teter T, Thoden D, Titov A, Tolbanov O,
  Torgrimsson G, Tyazhev A, Wing M, Zanetti M, Zarubin A, Zeil K, Zepf M and
  Zhemchukov A 2021  (arXiv:2102.02032)

\bibitem{AdvatechLYSO}
{Advatech - UK} 2021 {{LYSO(Ce) Scintillator Crystal}}
  \url{https://www.advatech-uk.co.uk/lyso_ce.html}, {Accessed}: 25 Jul 2021

\bibitem{Photek.2021}
{Photek} 2021 {{Image Intensifier User Guide}}
  \url{https://www.photek.com/pdf/user-guides/Photek--Image-Intensifier--UserGuide.pdf},
  {Accessed}: 20 Aug 2021

\bibitem{Orcas.2021}
{Hamamatsu} 2021 {{Technical note: ORCA-Flash4.0 V3 Digital CMOS camera
  C13440-20CU}}
  \url{https://www.hamamatsu.com/resources/pdf/sys/SCAS0134E_C13440-20CU_tec.pdf},
  {Accessed}: 20 Aug 2021

\bibitem{Bartoszek.1991}
Bartoszek L, Bharadwaj V, Church M~D, Hahn A~A, Peoples J, Pordes S~H, Rapidis
  P~A, Werkema S~J, Agahi D, Broemmelsiek D~R, Fast J~E, Gee M, Gollwitzer K~E,
  Mandelkern M~A, Marques J~L, Schultz J, Weber M~F, Ginsburg C~M, Masuzawa M,
  Ray R~E, Rosen J~L, Trokenheim S, Zhao J~L, Armstrong T~A, Hufford G~E, Lewis
  R~A, Majewska A~M, Reid J~D, Smith G~A, Hasan M~A, Biino C and Palestini S
  1991 {\em Nucl. Instrum. Methods Phys. Res. A\/} {\bf 301} ISSN 0168-9002

\bibitem{Kirsebom.1986}
Kirsebom K and Sollie R 1986 {\em Nucl. Instrum. Methods Phys. Res. A\/} {\bf
  245} ISSN 0168-9002

\bibitem{PMTdatasheet}
{Et}-{Enterprises} 2016 9111b series data sheet
  \url{https://et-enterprises.com/images/data_sheets/9111B.pdf}, {Accessed}: 22
  Mar 2021

\bibitem{Picoscope.2020}
Picotech 2021 {{PicoScope 6000 - high performance USB scope}}
  \url{https://www.picotech.com/oscilloscope/6000/picoscope-6000-overview},
  {Accessed}: 09 Mar2021

\bibitem{agostinelli_geant4simulation_2003}
Agostinelli S {\em et~al.\/} 2003 {\em Nucl. Instrum. Meth. A\/} {\bf 506}
  250--303

\bibitem{allison_geant4_2006}
Allison J {\em et~al.\/} 2006 {\em IEEE Trans. Nucl. Sci.\/} {\bf 53} 270--278

\bibitem{allison_recent_2016}
Allison J {\em et~al.\/} 2016 {\em Nucl. Instrum. Meth. A\/} {\bf 835} 186--225

\bibitem{FACETII.2019}
Yakimenko V, Alsberg L, Bong E, Bouchard G, Clarke C, Emma C, Green S, Hast C,
  Hogan M~J, Seabury J, Lipkowitz N, O'Shea B, Storey D, White G and Yocky G
  2019 {\em Phys. Rev. Accel. Beams\/} {\bf 22}(10) 101301

\bibitem{meuren_semiclassical_2016}
Meuren S, Keitel C~H and Di~Piazza A 2016 {\em Phys. Rev. D\/} {\bf 93}(8)
  085028

\bibitem{fluka_2005}
Fassò A, Ferrari A, Ranft J and Sala P~R 2005 {\em CERN-2005-10 (2005),
  INFN/TC\_05/11, SLAC-R-773\/}

\bibitem{bohlen_fluka_2014}
Böhlen T, Cerutti F, Chin M, Fassò A, Ferrari A, Ortega P, Mairani A, Sala P,
  Smirnov G and Vlachoudis V 2014 {\em Nucl. Data Sheets\/} {\bf 120} 211--214

\bibitem{Seifert.2012}
Seifert S, Steenbergen J~H~L, {van Dam} H~T and Schaart D~R 2012 {\em J.
  Instrum.\/} {\bf 7} ISSN 1748-0221

\bibitem{EBT3.2020}
Ashland 2021 {{Gafchromic Dosimetry Media, Type EBT-3, Ashland Inc.}}
  \url{http://www.gafchromic. com/documents/EBT3_Specifications.pdf},
  {Accessed}: 10 Apr 2021

\bibitem{Epson.2020}
Epson 2021 {{User’s Guide Perfection V700 Photo/V750 Pro, Seiko Epson K.K.}}
  \url{https://files.support. epson.com/pdf/prv7ph/prv7phug.pdf}, {Accessed}:
  10 Apr 2020

\bibitem{wulff_construction_2020}
Wulff J 2020 Construction and {Calibration} of a {Single} {Particle}
  {Cherenkov} {Calorimeter} for {Strong}-field {QED} {Experiments}, {Bachelor
  thesis}

\bibitem{WrattenFilter}
KODAK 2021 {KODAK Wratten 2 Filters}
  \url{https://www.kodak.com/en/motion/page/wratten-2-filters}, {Accessed}: 25
  Jul 2021

\bibitem{MAX_EBT_Calibration}
M\"{a}usezahl M 2019 {Untersuchung lasergetriebener Protonenbe-schleunigung
  bezüglich Vorplasmaerzeugungund räumlicher Protonendetektion}, {Master
  thesis},
  \url{https://www.rlp-ioq.uni-jena.de/ioqjpmedia/PDFs/2019_Maeusezahl_Masterarbeit.pdf},
  {Accessed}: 25 Jul 2021

\bibitem{Taylor.1997}
Taylor J~R 1997 {\em {An introduction to error analysis: The study of
  uncertainties in physical measurements}\/} 2nd ed ({University Science}) ISBN
  978-0935702750

\end{thebibliography}

\end{document}